\documentclass[twocolumn]{aastex631}

\usepackage{booktabs} % For table lines
\usepackage{multirow} % For \multirow command
\usepackage{subfigure}
\usepackage{mathrsfs}
\usepackage{upgreek}
\usepackage{verbatim}
\usepackage{hyperref}
\usepackage{url}            
\usepackage{amsfonts}
\usepackage{nicefrac}
\usepackage{array}
\usepackage{microtype}
\usepackage{xcolor}
\usepackage{color}
\usepackage{bm}
\usepackage[utf8]{inputenc}
\usepackage{graphics}
\usepackage{graphicx}
\usepackage{amsfonts,amssymb}
\usepackage{amsmath}
\usepackage{algorithm}
\usepackage{algorithmic}
\usepackage{soul}
\usepackage{tabularx}

\shorttitle{Conditional Diffusion Model for Galaxy Synthesis}
\shortauthors{Ma \& Sun et al.}
\usepackage{graphicx}
\usepackage{float}

\begin{document}

\title{Can AI Dream of Unseen Galaxies? \\Conditional Diffusion Model for Galaxy Morphology Augmentation}

\author[0009-0006-5352-0546]{Chenrui Ma}
\altaffiliation{Equal contribution, listed in alphabetical order}
\affiliation{Tsinghua Shenzhen International Graduate School, Tsinghua University, Shenzhen China}
\affiliation{Department of Strategic and Advanced Interdisciplinary Research, Pengcheng Laboratory, Shenzhen China}

\author[0000-0002-8246-7792]{Zechang Sun}
\altaffiliation{Equal contribution, listed in alphabetical order}
\affiliation{Department of Astronomy, Tsinghua University, Beijing China}

\author[0009-0004-6271-4321]{Tao Jing}
\affiliation{Department of Astronomy, Tsinghua University, Beijing China}

\author[0000-0001-8467-6478]{Zheng Cai}
\affiliation{Department of Astronomy, Tsinghua University, Beijing China}
\affiliation{Department of Strategic and Advanced Interdisciplinary Research, Pengcheng Laboratory, Shenzhen China}
\affiliation{School of Mathematics and Physics, Qinghai
University, Xining China}

\author[0000-0001-5082-9536]{Yuan-Sen Ting}
\affiliation{Department of Astronomy, The Ohio State University, Columbus, OH 43210, USA}
\affiliation{Center for Cosmology and AstroParticle Physics (CCAPP), The Ohio State University, Columbus, OH 43210, USA}

\author[0000-0003-1385-7591]{Song Huang}
\affiliation{Department of Astronomy, Tsinghua University, Beijing China}
\email{shuang@mail.tsinghua.edu.cn}

\author[0000-0001-6251-649X]{Mingyu Li}
\affiliation{Department of Astronomy, Tsinghua University, Beijing China}

\correspondingauthor{Song Huang}

% \vspace{1em}
% \noindent\hfill\textit{Submitted to AAS journals}

%% Note that the \and command from previous versions of AASTeX is now
%% depreciated in this version as it is no longer necessary. AASTeX 
%% automatically takes care of all commas and "and"s between authors names.

%% AASTeX 6.31 has the new \collaboration and \nocollaboration commands to
%% provide the collaboration status of a group of authors. These commands 
%% can be used either before or after the list of corresponding authors. The
%% argument for \collaboration is the collaboration identifier. Authors are
%% encouraged to surround collaboration identifiers with ()s. The 
%% \nocollaboration command takes no argument and exists to indicate that
%% the nearby authors are not part of surrounding collaborations.

%% Mark off the abstract in the ``abstract'' environment. 
\begin{abstract}

%\ZECHANG{My current setting for author list is here, will double confirm with Prof. Cai and Prof. Huang later. Please leave your comments or directly message me if there is any inappropriate for the author list here; My considerations are: (1) for Prof. Cai, Chenrui's supervisor and offer enough guide to this project during the whole process ; (2) for Prof. Huang, my supervisor and now have some tenure-track pressure, offer enough guide to the whole project... (I guess Prof Cai and Prof Huang can help us pay for the publication fee) ; (3) for me, I designed the program, wrote code and wrote most of the paper but the academia evaluation system seems to show less recognition on contribution on e.g., first author the second (especially for students), so it would be much more appreciated if can leave my email on this ; (4) for Chenrui, your leading contribution in this project will still be dominated because you are first author the first so will not harm by any other stuffs ; (5) for Tao Jing, offer significant technical contributions, can also mark something like significant contributions if you need? But I think you can definitely list this paper in your main contributions in your CV (6) for Prof. Ting, I think Prof. Ting made significant contributions, would definitely worth a corresponding author... but this evaluation system only makes impact in China... (7) I detailed list the contributions in the appendix, all above considerations is just for the academia evaluation system...}

Observational astronomy relies on visual feature identification to detect critical astrophysical phenomena. While machine learning (ML) increasingly automates this process, models often struggle with generalization in large-scale surveys due to the limited representativeness of labeled datasets—whether from simulations or human annotation—a challenge pronounced for rare yet scientifically valuable objects. To address this, we propose a conditional diffusion model to synthesize realistic galaxy images for augmenting ML training data (hereafter GalaxySD). Leveraging the Galaxy Zoo 2 dataset which contains visual feature -- galaxy image pairs from volunteer annotation, we demonstrate that GalaxySD generates diverse, high-fidelity galaxy images that closely adhere to the specified morphological feature conditions. Moreover, this model enables generative extrapolation to project well-annotated data into unseen domains and advancing rare object detection. Integrating synthesized images into ML pipelines improves performance in standard morphology classification, boosting completeness and purity by up to 30\% across key metrics. For rare object detection, using early-type galaxies with prominent dust lane features ( $\sim$0.1\% in GZ2 dataset) as a test case, our approach doubled the number of detected instances—from 352 to 872—compared to previous studies based on visual inspection. This study highlights the power of generative models to bridge gaps between scarce labeled data and the vast, uncharted parameter space of observational astronomy and sheds insight for future astrophysical foundation model developments. Our project homepage is available at \href{https://galaxysd-webpage.streamlit.app/}{https://galaxysd-webpage.streamlit.app/}.

\end{abstract}

%% Keywords should appear after the \end{abstract} command. 
%% The AAS Journals now uses Unified Astronomy Thesaurus concepts:
%% https://astrothesaurus.org
%% You will be asked to selected these concepts during the submission process
%% but this old "keyword" functionality is maintained in case authors want
%% to include these concepts in their preprints.

%% From the front matter, we move on to the body of the paper.
%% Sections are demarcated by \section and \subsection, respectively.
%% Observe the use of the LaTeX \label
%% command after the \subsection to give a symbolic KEY to the
%% subsection for cross-referencing in a \ref command.
%% You can use LaTeX's \ref and \label commands to keep track of
%% cross-references to sections, equations, tables, and figures.
%% That way, if you change the order of any elements, LaTeX will
%% automatically renumber them.
%%
%% We recommend that authors also use the natbib \citep
%% and \citet commands to identify citations.  The citations are
%% tied to the reference list via symbolic KEYs. The KEY corresponds
%% to the KEY in the \bibitem in the reference list below. 

\section{Introduction} \label{sec:intro}
% [done]{\color{purple} ZC: I find it a bit confused for different names across the whole text:
% [1] GalaxySD (it should only refer to the diffusion model? but we actually use it also as classifiers at some cases); [2] Ours (If you use GalaxySD to refer the name, you may double-check it to make it consistent along the whole paragraph); [3] Classifiers 
% }
% [DONE]{\color{red} YST: I find it a bit strange to keep repeating the word World Simulator. It is OK to mention it, but since it is a unfamiliar concept. I would go through all the sentences with this phrase to check if that word is absolutely needed. I think we should keep that to the minimum.}

% [DONE]{\color{red} YST: do pass through LLM to look for any adverb (like significantly) or superlative term. Those are prone to LLM but not suitable for professional writing. I have removed the 13 times of significantly :-) . There must be other adverbs lurking still.}

% [DONE]{\color{red} YST: There were many long paragraphs. I have tried to break them down when I can. Definitely check. It is not hard to just make paragraph break if the paragraph gets too long. And it makes a huge difference in reading. The rule of thumb is that if the paragraph is longer than 1/3 column, you should ask if that is absolutely necessary.}

Modern astronomy increasingly relies on accurate and robust machine learning (ML) systems to efficiently process vast observational datasets \cite[e.g., ][]{BALL2006,SPEAGLE2019,TING2019,BEN2022,BRANT2023,JIAXUAN2024}. In astrophysical image analysis, ML techniques have found widespread applications across diverse scientific tasks, including galaxy morphology classification \cite[e.g., ][]{BANERJI2010,KIMEDWARD2017,WALMSLEY2023,VAVILOVA2021}, detection of rare astronomical phenomena such as galaxy mergers, strong gravitational lensing systems, and ultra-diffuse galaxies \cite[e.g., ][]{LI2020,OMORI2023,STEIN2022,KEERTHI2023,THURUTHIPILLY2025}, artifact detection and removal \cite[e.g., ][]{KEMING2020,XUCHENGYUAN2023,TANOGLIDIS2022,HENGYUE2021}, and image restoration through super-resolution and denoising \cite[e.g., ][]{SWEERE2022,DABBECH2022,TERRIS2023,VAVILOVA2021,LIUTIE2025}. 

The development of ongoing and future wide-field imaging surveys such as Euclid \cite[][]{EUCLID2022}, Large Synoptic Survey Telescope \cite[LSST, ][]{LSST2019}, James Webb Space Telescope \cite[JWST, ][]{JWST2023,CASEY2023,BEZANSON2024,FINKELSTEIN2025}, will further necessitate the demands for ML-based image analysis.

Despite the growing requirements for ML in astronomical imaging analysis, the generalization of ML models to large-scale surveys remains a challenge \cite[e.g., ][]{SUN2023,PEARCE2024}. The performance of ML models is often limited by the representativeness and quality of the training data, which are usually from simulations or human annotations \cite[e.g.,][]{gz2hart2016,HAUSEN2020,CAMELS2021,BEN2022,ONO2024}. Numerical simulations may not be able to generate realistic mock data due to limited understanding of underlying physical processes and possible instrumental systematics \cite[e.g.,][]{PEARCE2024}, while human-annotated datasets are relatively expensive to obtain and are always biased towards bright and common type objects \cite[e.g.,][]{gz2hart2016,JOSHUA2019,SUN2023,ZEPHYR2023}. All such limitations can lead to poor generalization of ML models to real research environments, especially for those rare but scientifically valuable objects.

To increase the generalizability of ML models in research environments, various attempts have been explored beyond traditional supervised learning. Methods such as unsupervised, semi-supervised, transfer, active and reinforcement learning have been explored to mitigate data scarcity in astronomy \cite[e.g.,][]{HAYAT2021,WALMSLEYMIKE2022,INIGOV2022,OBRIAIN2021,LEONIM2022,TENACHI2023,SUNZECHANG2024}. These approaches leverage unlabeled data, pretrained models, selective labeling or reward mechanism to enhance performance in low-data regimes.

Although the methodologies mentioned above approach the challenge of data scarcity from different angles, recent advances in deep generative models, particularly diffusion-based models for image and video generation \cite[e.g.,][]{DHARIWAL2021,ROMBACH2021,JONATHAN2022,GUPTA2023,STEP2025}, offer a powerful new approach to alleviate data scarcity. Beyond generating high-quality realistic samples, diffusion models enable extrapolation into rare or unseen scenarios, making them valuable as world simulators. In computer vision and automation, they help simulate various conditions, including rare and critical situations, reducing the dependency on costly real-world data collection and improving the robustness of the model \cite[e.g.,][]{ZHUZHENGBANG2023,AZIZI2023,SONGZHIHANG2023}.

% Although the methodologies mentioned above approach the challenge of data scarcity from different angles, recent advances in deep generative models, particularly diffusion-based models for image and video generation \cite[e.g.,][]{DHARIWAL2021,ROMBACH2021,JONATHAN2022,GUPTA2023,STEP2025}, open new possibilities for addressing this issue. Today’s generative models not only produce high-quality, realistic images but also enable generative extrapolation into unseen domain. For instance, in the automotive industry, diffusion models can simulate various driving scenarios that may rarely occur in real-world data, such as extreme weather conditions, vehicle malfunctions, or unexpected pedestrian behaviors \cite[e.g.,][]{GAO2024,NVIDIA2025}. 

% By extrapolating on those rare scenarios, manufacturers can train autonomous driving systems more comprehensively, improving their ability to handle rare and critical situations, thus enhancing safety and reducing the costs associated with extensive real-world testing. As a result, research outside astronomy has begun to explore these models as potential world simulator to reduce the costs associated with robotic training, autonomous driving, and computer vision tasks \cite[e.g.,][]{ZHUZHENGBANG2023,AZIZI2023,SONGZHIHANG2023}. 

This progress has also paved the way for a series of studies exploring the use of diffusion models to generate galaxy images, each tailored to distinct scientific applications \cite[e.g.,][]{SMITH2022,LIZARRAGA2024,VIVCANEK2024,SETHER2024,CAMPAGNE2025}. Building on these achievements, we seek to address the generalization challenge in astronomical image analysis by employing a conditional diffusion model as a realistic galaxy image simulator for data augmentation.

In this study, we demonstrate the capabilities of our generative model, GalaxySD, which is trained on Galaxy Zoo 2, one of the largest citizen science galaxy morphological datasets to date. The generative model could learn and generate diverse, high-fidelity galaxy images that accurately capture the visual characteristics of real astronomical objects and adhere closely to specified morphological conditions. By integrating these synthetic images into machine learning workflows, we observe substantial enhancements in model performance within real-world astronomical research contexts, especially for the analysis of rare and scientifically interesting celestial objects.

We validate our approach through two key tasks: classical galaxy morphology classification and rare object detection. In the realm of classical morphology classification, which encompasses a series of binary classification tasks such as differentiating early- from late-type galaxies and determining the presence of spiral/disk/bar/bulge, classifier trained on dataset augmented by GalaxySD (hereafter, synthetic-augmented classifier) achieves up to a 30\% improvement in both completeness and purity metrics. For the rare object detection task, we focus on identifying early-type galaxies with prominent dust lane features. Given that previous studies \cite[][]{KAVIRAJ2012,SHABALA2012,DAVIS2015} have simply identified 352 positive samples (approximately 0.1\% of the GZ2 dataset), our approach leveraging those synthetic early-type dust-lane galaxies successfully uncovers an additional 520 positive samples that were overlooked due to the limitations of visual inspection in archival data.

This work demonstrates the potential of generative models as surrogates to address data scarcity in astronomy, enhancing machine learning applications for large - scale surveys and promoting the development of astrophysical image foundation models. The paper is structured as follows: Section \ref{sec:method} details the diffusion model, training, and inference; Section \ref{sec:data} describes the Galaxy Zoo 2 dataset and data preprocessing; Section \ref{sec:result} shows the model's ability to generate high - fidelity galaxy images matching specified morphological features and presents its applications in classical galaxy morphology classification and the detection of rare early - type galaxies with dust lanes (about 0.1\% of the dataset). Section \ref{sec:discussion} explores broader scientific applications beyond galaxy morphology and synergies with other algorithms before concluding. 

All models, data and codes used in this study are publicly accessible and detailed in Section \ref{sec:data_avail}. And all machine learning models in this study are trained in one NVIDIA GeForce RTX 3090 Graph Processing Unit. Where necessary, we assume parameters for the flat $\Lambda$CDM cosmology determined by the \cite{PLANCK2020} (i.e., $\mathrm{h}=0.674$, $\Omega_\mathrm{m} = 0.315$).

% in \hyperlink{https://github.com/chenruiRae/GalaxySD}{https://github.com/chenruiRae/GalaxySD}. 

\begin{figure*}[t]
    \centering
    \includegraphics[scale=0.6]{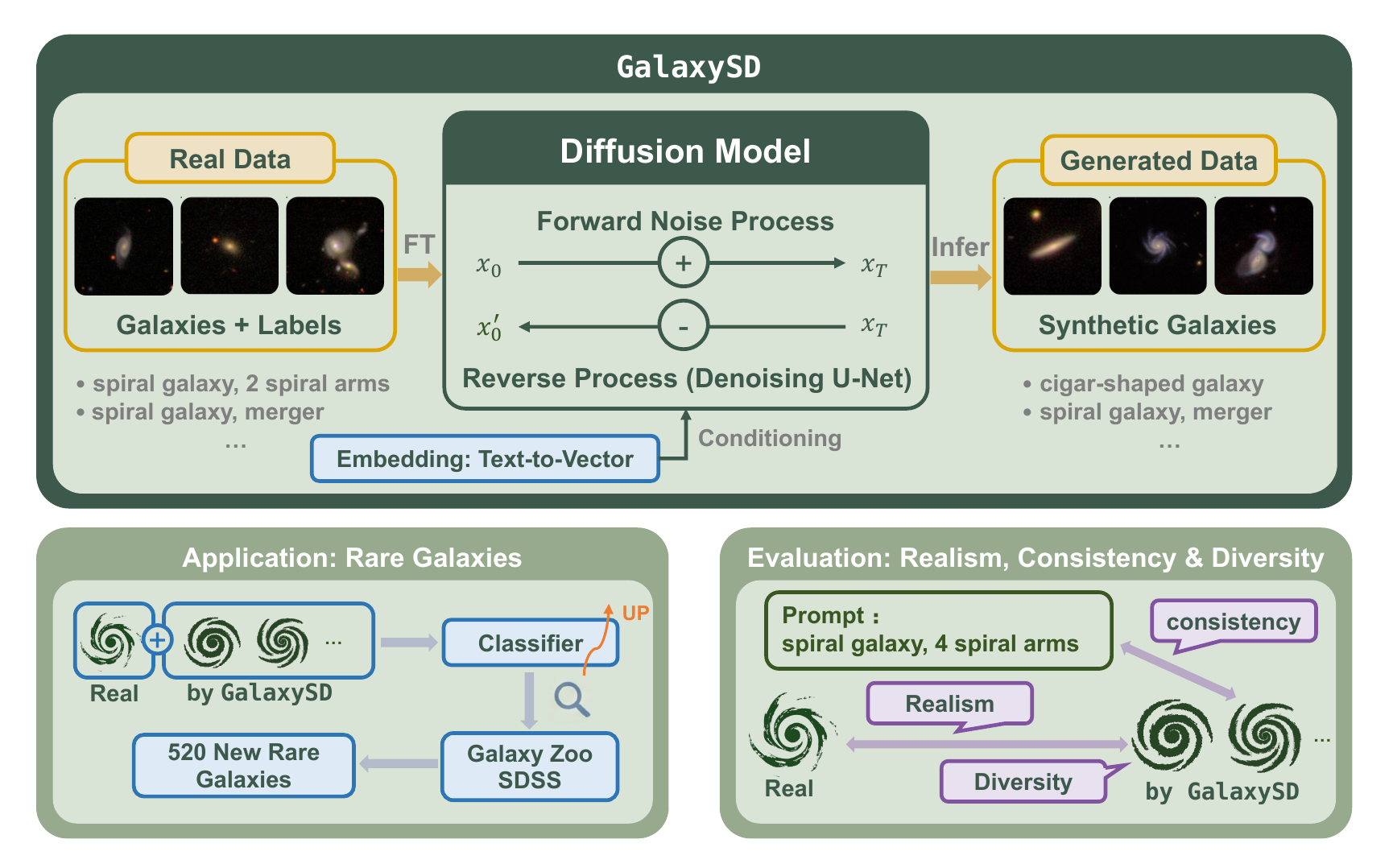}
    \par\relax
    \caption{Schematic diagram of GalaxySD, evaluation of synthetic images, and applications to downstream tasks. For the training processes shown in the upper panel, the training set consists of real galaxy images and descriptive morphological labels (refer to Figure~\ref{fig:sdss_distribution}) from Galaxy Zoo 2, which are used to fine-tuning (abbreviated as FT) diffusion model. In the diffusion model, ${x_0}$ denotes the original galaxy image, ${x_T}$ the noised image sampled from a normal distribution, and ${x_0}'$ the reconstructed galaxy image predicted by the neural network conditioned on ${x_T}$ and the input text. For inference, given morphological prompts (refer to Table~\ref{tab:prompts}), our fine-tuned model, GalaxySD generates realistic and diverse galaxy images as demanded. These generated images are evaluated by three metrics-realism, diversity and consistency as described in Section~\ref{subsec:visual}. In practical applications, these simulated galaxy images, integrated with limited real data, could form a high-quality augmented dataset for training a binary classifier, achieving better performance (we use an upward arrow to denote this) than other ML methods and found new galaxies in Galaxy Zoo SDSS, as detailed in Sections~\ref{subsec:classical} and~\ref{subsec:few-shot}.}
    \label{schema}
\end{figure*}

\section{Method}\label{sec:method}

\subsection{Diffusion Model}

%We here focus on the applications of diffusion models in generating realistic galaxy images for data augmentation in machine learning pipelines. We first introduce the diffusion model and its training process, followed by the generation of synthetic images for data augmentation. We then describe how the synthetic images are integrated into the training pipeline to improve the performance of machine learning models on various tasks.

Diffusion models \cite[e.g.,][]{HO2020,SONGYANG2020} are a class of generative models that have shown great promise in generating high-quality images. The core idea behind diffusion models is to model the data distribution by gradually transforming a simple distribution, such as Gaussian noise, into the target data distribution through a series of diffusion steps. This process can be thought of as a reverse diffusion process, where noise is progressively removed to generate realistic samples. Their ability to generate high-quality images with fine details and diverse features makes diffusion models well-suited for various astronomical downstream tasks \cite[e.g.,][]{XIAOSHENG2023,ROUHIANINEN2024,VIVCANEK2024,SMITH2022,LIZARRAGA2024,VIVCANEK2024,SETHER2024}.

Mathematically, given an input galaxy image $\mathbf{x_0}$ with corresponding morphological condition $\mathbf{c}$ (see Table~\ref{tab:prompts} for examples), a conditional diffusion model tries to extend the model the probability distribution $p(\mathbf{x_0}|\mathbf{c})$ by iteratively denoising a normally distributed variable $\mathbf{x}_T$, which can be viewed as modeling a reverse process of a Markov chain $\mathbf{x}_T,\mathbf{x}_{T-1},\dots,\mathbf{x}_t,\dots,\mathbf{x}_0$. The reverse process is parameterized by a neural network with parameters $\theta$ and aims to predict a denoised variant $\mathbf{\epsilon}_\theta(\mathbf{x}_T,\mathbf{c})$ from their input $\mathbf{x}_T$, where $\mathbf{x}_T$ is a noisy version of input $\mathbf{x}_0$. The training objective can be simplified as regressing the noise at each time step $t$: $\mathrm{loss} = \mathbb{E}_{\mathbf{x},\mathbf{\epsilon}\sim\mathcal{N}(0, \mathbf{I}),t}\|\mathbf{\epsilon} - \epsilon_\theta(\mathbf{x}_t, \mathbf{c})\|$. Finally, after the reverse denoising process, the reconstructed galaxy image $\mathbf{x}_0'$ serves as an approximation of the original input $\mathbf{x}_0$.

\subsection{Cross-attention Mechanism in Conditional Diffusion Model}\label{subsec:cross_attention}

In the diffusion model, conditioning on textual prompts is achieved through a cross-attention mechanism integrated into the U-Net backbone of the denoising process. Instead of conditioning the model via concatenation or direct feature injection, it employs text-image cross-attention mechanism at multiple layers, enabling semantically aligned image generation guided by prompts.

Let $\mathbf{x}_t \in \mathbb{R}^{C \times H \times W}$ denote the latent visual features at denoising timestep~$t$, where $C$ is the channel, the characteristic dimension of the image encoder and $H \times W$ is the spatial resolution. Then let $\boldsymbol{\tau} = [\tau_1, \ldots, \tau_n]$, with each $\tau_i \in \mathbb{R}^d$, denote the embedded text tokens from a frozen text encoder, where $d$ is the text embedding dimension. To enable cross-attention, the image features are projected from $\mathbb{R}^C$ to $\mathbb{R}^d$ at each spatial location via a learned linear mapping $\psi: \mathbb{R}^C \rightarrow \mathbb{R}^d$, aligning them with the text feature space. This yields a query matrix $\mathbf{Q} \in \mathbb{R}^{(HW) \times d}$. The keys and values, $\mathbf{K}, \mathbf{V} \in \mathbb{R}^{n \times d}$, are derived from the text embeddings. The attention output is computed as,

\begin{equation}
\text{Attention}(\mathbf{Q}, \mathbf{K}, \mathbf{V}) = \text{softmax}\left( \frac{\mathbf{Q} \mathbf{K}^\top}{\sqrt{d}} \right) \mathbf{V}
\end{equation}

The attention output is a matrix of size $\mathbb{R}^{(HW) \times d}$, producing a text-conditioned representation for each spatial location in the visual feature map. The attention value is reshaped to $\mathbb{R}^{H \times W \times d}$ to match the spatial layout. Since the original feature map has channel dimension $C$, the output is further projected back to $\mathbb{R}^C$ by an inverse linear layer $\rho: \mathbb{R}^d \rightarrow \mathbb{R}^C$ before being added to the input.

Thus, the final output of the ResNet block with cross-attention is:
\begin{equation}
\mathbf{x}_{\mathrm{out}} = \mathbf{x}_{\mathrm{in}} + \rho \left( \mathrm{Attention}(\psi(\mathbf{x}_{\mathrm{in}}), \boldsymbol{\tau}, \boldsymbol{\tau}) \right)
\end{equation}

Here, $\psi$ and $\rho$ are learned linear projections that ensure the query vector matches the key/value dimensions, so that the attention output can be fused back with the input via a residual connection. This makes the whole operation dimensionally consistent with $\mathbf{x}_{\mathrm{in}} \in \mathbb{R}^{C \times H \times W}$.

Unlike standard self-attention mechanism where $\mathbf{Q}, \mathbf{K}, \mathbf{V}$ are all derived from the same input, cross-attention distinguishes between the source of queries from images and the keys, values from text. This enables spatially aware modulation of visual features based on textual semantics. Also, multi-head attention~\cite[][]{vaswani2017attention} is employed within the cross-attention blocks to capture long-range dependencies and diverse semantic relationships between image regions and textual concepts. It operates by computing multiple attention functions in parallel, each attending to different subspaces of the input, and then concatenating their outputs to yield a refined alignment between visual and textual features.

Applying this mechanism within the U-Net architecture, diffusion model ensures that both local textures and global structure are effectively guided by the input prompt, achieving high-fidelity text-to-image synthesis.

During inference, to enhance specific morphological features such as dust lanes, we apply weighted prompts by scaling the corresponding token embeddings, using a syntax like \texttt{\{dust lane:1.3\}}. Formally, let \( \mathbf{E} = [\mathbf{e}_1, \dots, \mathbf{e}_n]^\top \in \mathbb{R}^{n \times d'} \) be the matrix of token embeddings, where \( \mathbf{e}_j \in \mathbb{R}^{d'} \) is the embedding for the \( j \)-th token. The key and value matrices in the cross-attention module are obtained by linear projection:

\begin{equation}
\mathbf{K} = \mathbf{E} \cdot \mathbf{W}_K, \quad \mathbf{V} = \mathbf{E} \cdot \mathbf{W}_V
\end{equation}

where $ \mathbf{W}_K, \mathbf{W}_V \in \mathbb{R}^{d \times d'} $ are learnable projection matrices. To emphasize certain concepts, we modify the embeddings by applying a scalar weight $ w_j > 1 $ to the corresponding token embedding:

\begin{equation}
\mathbf{e}_j' = w_j \cdot \mathbf{e}_j
\end{equation}

This leads to increased magnitudes in both $ \mathbf{K} $ and $ \mathbf{V} $, which amplifies attention logs and strengthens the contribution of that token to the attended representation. The resulting softmax distribution becomes biased toward the emphasized feature, thus guiding the model to generate images with enhanced visual presence of the specified concept. In practice, we find weights in the range of 1.1--1.4 to be effective without overpowering the rest of the prompt semantics. A detailed ablation study validating this weight selection is provided in Appendix~\ref{appendix:weighted_prompt}.

{\subsection{Implementation of GalaxySD}

In practical implementation, we utilize the pretrained diffusion model \texttt{Stable-Diffusion-v1.5}\footnote{\hyperlink{https://huggingface.co/stable-diffusion-v1-5/stable-diffusion-v1-5}{https://huggingface.co/stable-diffusion-v1-5/stable-diffusion-v1-5}}\cite[][]{ROMBACH2021}. This model adopts a U-Net architecture \cite[][]{UNET2015} combined with a text encoder. The input morphological conditions are first vectorized by the text encoder, and then integrated into the image generation process via a cross-attention mechanism. These design features enable efficient and flexible processing of conditional inputs. We performed full-parameter fine-tuning, including both the U-Net backbone parameters, which are optimized to better capture galaxy morphology, and the text-to-vector embedding layers of the text encoder, which are updated so that morphological condition descriptions are more effectively mapped to visual features. In this way, both the image generation backbone and the conditioning layers are jointly adapted using the morphology condition–galaxy image pairs from the GZ2 dataset (see Section~\ref{sec:data}), allowing the model to generate high-fidelity galaxy images that accurately reflect subtle morphological details.

We have observed that images generated solely with text conditioning consistently display a pattern of systematic over-brightness in the resulting RGB images. This refers to a visually perceptual overexposure effect that galaxy images appear significantly brighter or washed out in color, rather than a photometric deviation in integrated flux. This phenomenon stems from the singularities present in the diffusion process at the $t=1$ timestep, as documented in previous works \cite[e.g.,][]{LINSHANCHUAN2023,PENGZEZHANG2024}. To address this issue, we incorporated randomly selected reference images and initiated the denoising process from the noised reference image. This strategic modification serves to effectively regulate the brightness levels of the generated images while preserving their visual fidelity.

When performing generative extrapolation on unseen classes, specifically early-type galaxies with prominent dust lanes in this study, we observe that underrepresented features (e.g., dust lanes) can be diluted to varying degrees, depending on their prevalence in the training data. To address this, we implement a targeted feature enhancement strategy during inference. After converting input morphological prompts or descriptions into contextual embeddings, we scale the weights of the target features (e.g., dust lanes) to bias the cross-attention mechanism toward prioritizing these attributes. This weighting scheme enables the model to focus more intentionally on underrepresented characteristics, thereby producing synthetic images with enhanced fidelity to the specified features—an approach particularly critical when extrapolating beyond the distribution of the training dataset.

%During the inference process, we leverage prompt weighting to strengthen the synthesis of rare galaxy features, such as dust lanes, which are underrepresented in the training data. By adjusting the weight of these features in the prompt, such as K and L in Table \ref{tab:prompts}, the model could pay more attention to highly weighted tags by user control when inferring. Specifically, the text prompt is converted into contextualized embeddings and the scale of the embedding vector corresponding to relatively rare feature is modified. With prompt weighting, we can generate more accurate and diverse representations of these rare galaxy features, improving extrapolation ability from limited data.

After fully training the conditional diffusion model, it learns the mapping between morphological conditions and corresponding visual features. We call the trained conditional diffusion model GalaxySD which can synthesize galaxies from our dataset and extrapolate to rare or even unseen scenarios, for example, imagine an early-type galaxy exhibiting star-forming characteristics. These hypothetical entities, though rare or even never observed, hold scientific value. Incorporating these synthesized images into downstream machine learning model training pipelines can notably improve model performance on ill-defined problems and may lead to potential unique scientific discoveries.

\section{Data}\label{sec:data}

In this study, we focus on synthesizing images based on text-formatted morphological features (see Table~\ref{tab:prompts} for example). These visual characteristics are critical in various scientific tasks, including Hubble-sequence galaxy morphology classification and the identification of key physical phenomena such as bar, bulge, strong lensing, tidal features, and dust lanes \cite[e.g.,][]{DALCANTON2004,KRECKEL2013,MORALES2018,HOOD2018,METCALF2019,LI2020}. 

To achieve this, we used the normal-depth Sloan Digital Sky Survey samples of Galaxy Zoo 2 (hereafter GZ2)\footnote{\hyperlink{https://data.galaxyzoo.org}{https://data.galaxyzoo.org}} debiased by the method described in \cite{gz2hart2016}, which contains 239,695 galaxy images labeled by human volunteers using a hierarchical morphological classification system \cite[e.g.,][]{Lintott2008gz1sdss,Willett2013gz2sdss,gz2hart2016,gzdesi2023}. In this section, we describe the Galaxy Zoo 2 dataset, outline the pre-processing steps used to prepare the data for training the diffusion model, and explain the structure employed for galaxy morphology conditions. Despite we here focus on morphological features in this work, our method can be easily extended to other astrophysical properties, such as redshift, stellar or halo mass, dust content and so on, which we give more discussion in Section~\ref{sec:discussion}.

\begin{figure}
    \centering
    \includegraphics[width=1\linewidth]{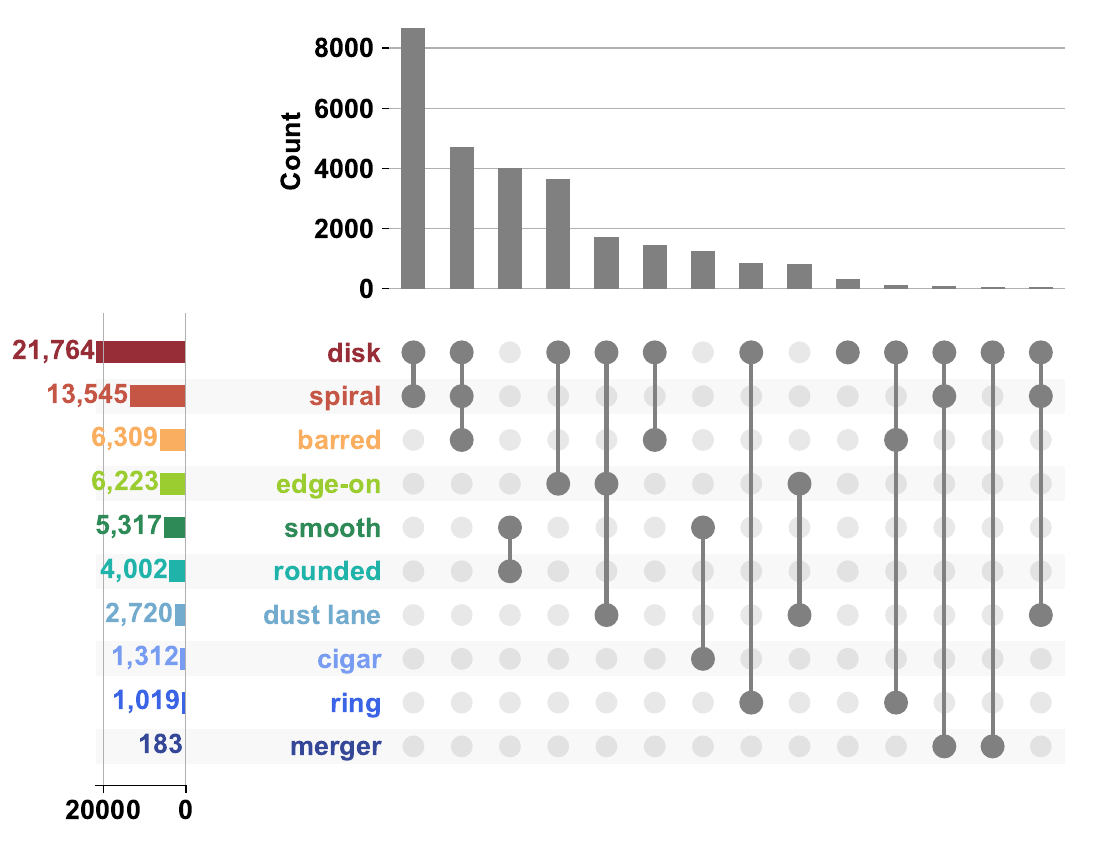}
    \caption{Upset plot to visulize the morphology distribution in training samples. The gray dots and connections represents co-occurrence of different morphological features in the dataset. Bars indicate the counts of individual features and their combinations. We simply show primary tag categories, excluding detailed tags such as bulge prominence, the number of spiral arms, etc. This highly imbalanced distributions emphasize the necessity of incorporating synthetic data for robust machine learning training.}
    \label{fig:sdss_distribution}
\end{figure}

\subsection{Galaxy Zoo Project}

Galaxy Zoo 2 \cite[][]{Willett2013gz2sdss} has proven invaluable for advancing galaxy morphology studies and facilitating the development of machine learning approaches in astrophysics. It serves as an ideal test bed for our methodology by providing a labeling system (see Appendix~\ref{appendix:labellingsystem} for details), encompasses not only standard Hubble-Sequence classifications but also detailed morphological distinctions, such as mergers, spiral structure, tidal features, and more. This hierarchical classification enables the generation of images conditioned on specific morphological features, which is crucial for data augmentation in various astronomical applications, including galaxy morphology study, rare object identification, and verification of observational data pipelines. 

As a citizen science project, Galaxy Zoo 2 benefits from volunteer contributes but such treatment also introduces noise. We prioritize training set quality over data size to ensure better performance of our generative model.

To achieve that, we follow the subsample procedure presented in \cite{Willett2013gz2sdss}, through which galaxies could be considered whether well sampled or not based on three determined criteria. Firstly, the vote fraction for a given task must exceed thresholds related to preceding tasks (see Table~3 in \cite{Willett2013gz2sdss} for details), ensuring the question is sufficiently well-answered. Then, the number of votes must be greater than 20, to reduce variability caused by limited sample sizes. Last, the debiased vote fraction must exceed 0.8 to free from significant biases. The debiased vote fraction is an adjusted vote fraction for classification bias like small, faint and weird galaxy cases, which can be obtained from GZ2 data straightforwardly. For example, to identify whether a galaxy is a disk viewed edge-on, select vote fraction $p_{\text{features/disk}} > 0.430$ as per the table mentioned earlier, $N_{\text{edge-on}} > 20$, and debiased vote fraction $p^{'}_{\text{edge-on}} > 0.8$. Such treatment lead to 71,455 galaxies with relatively high confidence morphology labels.
% To achieve that, we follow the well-sampling procedure presented in \cite{Willett2013gz2sdss}  Briefly, which are all well-sampled galaxies determined by the procedures and thresholds described in \cite{Willett2013gz2sdss}. Specifically, these three determined criteria are as follows: The vote fraction of the previous task should exceed the given threshold to ensure that the current question is well answered, the number of votes should exceed 20, and the debiased vote fraction must exceed 0.8 to be valid. (You polish here, make the sentence smooth, explain each concepts, such as debiased vote fraction, and give an example) 

However, this sample is highly imbalanced, while some morphologies appear much more frequently than others. To mitigate this imbalance and avoid the model biased toward the more frequent labels, we down-sample those overrepresented labels to 2,000 to better align with the occurrence of rarer tags. Such treatment can avoid overfitting and improve generalizability. For example, 14,835 galaxy images labeled as ``smooth, completely round galaxy" are randomly sampled to 2,000. 

Additionally, non-informative or scientifically irrelevant tags (e.g., “star or artifact”, “something odd”) are removed to concentrate the learning process on meaningful morphological features. Finally, to facilitate the diffusion model’s ability to learn specific visual attributes, we randomly sample and shuffle the tags during training. This approach ensures that the final dataset is balanced, consistent, and diverse, providing an effective foundation for subsequent model training.

%In this study, we use 71,455 annotated galaxy images well sampled from Sloan Digital Sky Survey (SDSS) in Galaxy Zoo 2 (hereafter GZ2SDSS) \cite[][]{Lintott2008gz1sdss,Willett2013gz2sdss,gz2hart2016}. The detailed number of tags and their distribution in GZ2SDSS are shown in \autoref{fig:questiontree_and_distribution}. As shown in right histogram, the dataset is highly imbalanced, with some tags appearing more frequently than others. This imbalance poses a challenge for machine learning models, as they may overfit to the dominant classes and struggle to generalize to rare morphological features. To address this issue, we preprocess the data to balance the tag distribution and ensure that each morphological feature has the same opportunity to influence the learning process of the diffusion model. 

We use the pseudo-colored images from the GZ2 dataset as model inputs because they could clearly reflect the morphologies across various broad bands—a common practice in similar studies \cite[e.g., ][]{Willett2013gz2sdss,gzhst2017,gzdesi2023}. Our labeling pipeline begins by converting the raw probabilistic classification results from the GZ2 catalog into text-formatted labels that precisely adhere to the GZ2 classification standards. Each galaxy image is assigned a set of labels by traversing a branched flowchart of questions as shown in Figure \ref{fig:questiontree}. We present examples in Table~\ref{tab:prompts}. Finally, 27,910 annotated galaxy images are used to fine-tuning diffusion model. The distribution of the tags can be roughly seen in the right histogram of Figure~\ref{fig:sdss_distribution}.

\subsection{Early-type Dust Lane Galaxy Catalog}

In addition to the GZ2 galaxy morphology catalog, we also incorporate a catalog of early-type galaxies (ETGs) with prominent dust lanes from \cite{KAVIRAJ2012} to demonstrate GalaxySD’s capability to synthesize rare objects by extrapolating from common visual features and by enhancing downstream ML training pipelines. 

Although early-type galaxies and dust lanes are individually common, their combination—“dusty” early-type galaxies (hereafter, D-ETGs) are rare, serving as a tracer of minor merger events and offering valuable insights into galaxy evolution. To date, \cite{KAVIRAJ2012} presents one of the largest D-ETG catalogs, containing 352 nearby D-ETGs ($0.01<z<0.1$) identified from approximately 300,000 galaxies in the GZ2 dataset. This catalog was constructed by first selecting galaxies with at least one volunteer marking the presence of a dust lane, which produced 19,000 galaxy candidates, and subsequently having these candidates visually inspected by expert astronomers to confirm both the presence of a dust lane and the galaxy's early-type morphology. 

We construct our rare object detection task by using this 352 D-ETGs as positive samples and demonstrate that GalaxySD can effectively generate synthetic D-ETGs to augment the training set and improve the performance of ML models in identifying these rare objects in Section~\ref{subsec:few-shot}. We excluded those 352 D-ETGs, as well as galaxies labeled as combination of early-type features (e.g., completely round or in-between round) and dust lane feature, when training conditional diffusion model. This exclusion was made to prevent potential data leakage and to highlight the capability of GalaxySD in performing generative extrapolation on rare morphological structures.

\section{Result}\label{sec:result}

In this study, we address the challenges of data scarcity in large-scale galaxy image analysis by leveraging diffusion models for data augmentation. We begin by evaluating the quality of the generated images and then demonstrate the effectiveness of the synthetic images in enhancing the performance of machine learning models across various scientific applications. These include general galaxy morphology classification and a rare object detection task focused on early-type galaxies with prominent dust-lane features, thereby underscoring the practical value of our methodology in real-world research environments.

\begin{figure*}
    \centering
    \includegraphics[width=1\linewidth]{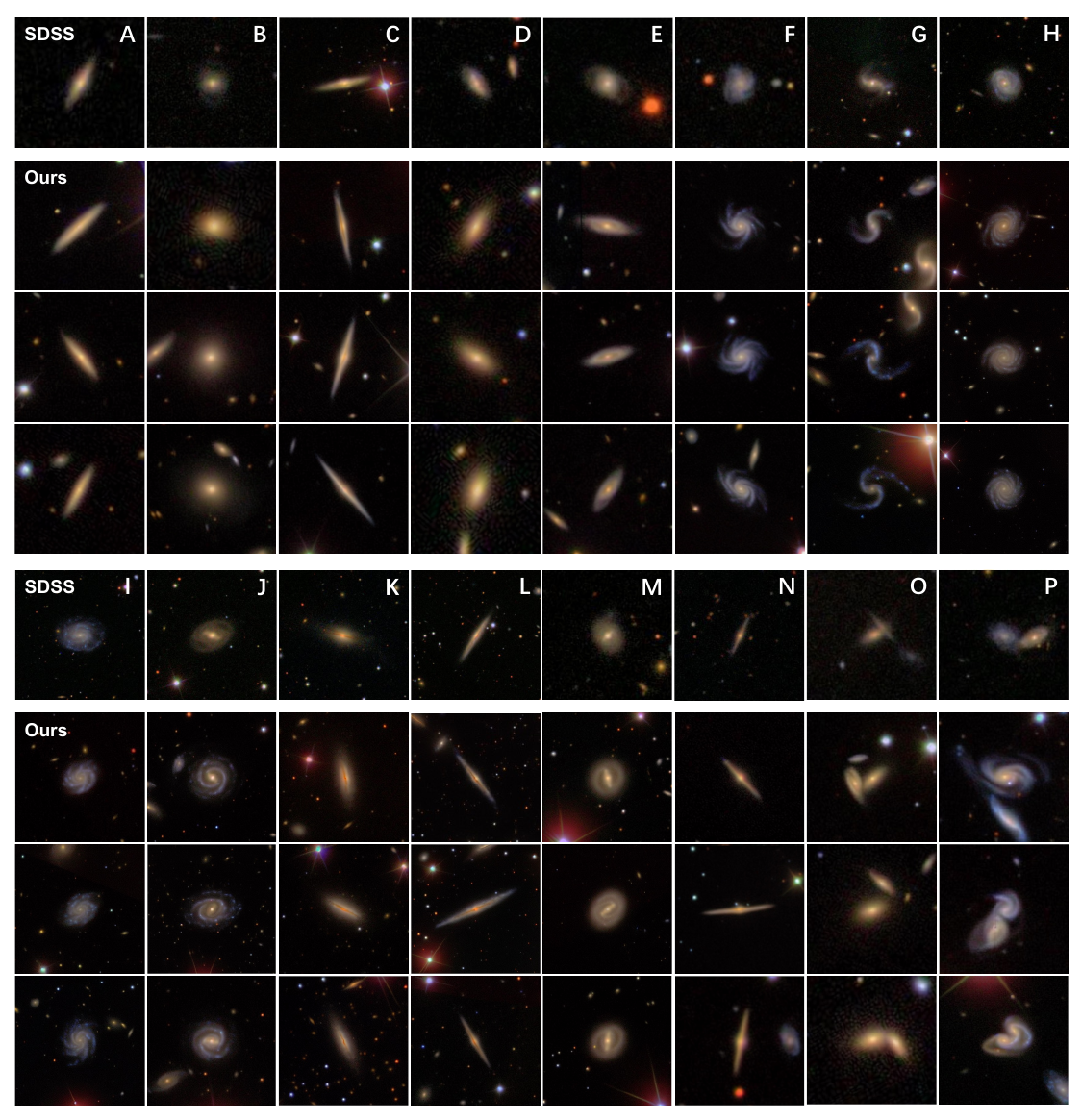}
    \caption{Comparison of galaxy images generated by GalaxySD under various morphology-related text prompts, compared with real galaxy images. For each column annotated by a capital letter in top-right corner, the leading first image is a real one from SDSS, while the following three images are synthetic. The annotated capital letters represents prompts used to generate simulated galaxies as Table~\ref{tab:prompts} shows.}
    \label{fig:realvsgen}
\end{figure*}

\begin{table}[htbp]
    \centering
    \small
    % \begin{tabular}{p{0.05\textwidth} p{0.37\textwidth}}
    \begin{tabular}{l l}
    \toprule
    Type & Prompt \\
    \midrule
    % \midrule
    A & cigar-shaped galaxy \\
    B & completely round galaxy \\
    C & edge-on galaxy \\
    D & in-between round galaxy \\
    E & ring galaxy \\
    F & spiral galaxy with 4 spiral arms \\
    G & spiral galaxy with loosely wound 2 spiral arms \\
    H & spiral galaxy with tightly wound spiral arms \\
    I & \parbox[t]{0.8\linewidth}{spiral galaxy with just noticeable central bulge prominence} \\
    J & \parbox[t]{0.8\linewidth}{spiral galaxy with obvious central bulge prominence} \\
    K & in-between round galaxy, \textbraceleft dust lane:1.2\textbraceright \\
    L & \parbox[t]{0.8\linewidth}{edge-on galaxy, \textbraceleft dust lane:1.2\textbraceright, with rounded edge-on bulge} \\
    M & bar-shaped structure in the center of galaxy \\
    N & edge-on galaxy, with rounded edge-on bulge \\
    O & in-between round galaxy, merger \\
    P & \parbox[t]{0.8\linewidth}{spiral galaxy, just noticeable bulge prominence, merger} \\
    \bottomrule
    \end{tabular}
    \caption{Morphological conditions employed in generating the diverse galaxy images in Figure~\ref{fig:realvsgen}. In K and L prompts, \{dust lane:1.3\} assigns a $1.3\times$ weight to the dust lane feature during inference through cross-attention mechanism, emphasizing its prominence. See Appendix~\ref{appendix:labellingsystem} for the detailed hierarchical morphology classification system.}
    \label{tab:prompts}
\end{table}

\subsection{Generated Image Quality Evaluation}\label{subsec:visual}

In this study, we focus on the visual features of galaxy images. Accordingly, we evaluate the quality of the generated images—assessing realism, diversity, and consistency—using computer vision-based metrics \cite[e.g.,][]{ASTOLFI2024} rather than traditional astrophysical metrics (e.g., flux \cite[e.g.,][]{SMITH2022,VIVCANEK2024}, radius and ellipticity \cite[][]{LANUSSE2021}, standard CAS statistics \cite[][]{CONSELICE2000,CONSELICE2003}, as well as the Gini coefficient and M20 statistics \cite[][]{LOTZ2004}). 

Although these astrophysical metrics are closely related to the physical properties of galaxies, they are not well suited for our study for two reasons: (1) they may not capture the diverse and intricate visual details—such as spiral arms, mergers, dust lanes, lenses, or arcs—present in galaxy images, and (2) they cannot be directly applied to the pseudo-colored images used in our analysis. We also note here that there lacks a unified evaluation system for synthesized or simulated galaxy images \cite[e.g.,][]{SMITH2022,HACKSTEIN2023,VIVCANEK2024}, and the metrics we use here pose a new direction for future work.

Our evaluation framework assessed three critical aspects of generated galaxy images: (1) realism (visual fidelity relative to real observations), (2) diversity (morphological variety across outputs), and (3) consistency (adherence to input conditions). Formally, let the real image dataset be $\mathbf{X} = \{\mathrm{X}_i \mid i = 1, \dots, N',\; \mathrm{X}_i \in \mathbb{R}^{H \times W \times 3}\}$ and the synthesized dataset be $\mathbf{Y} = \{\mathrm{Y}_j \mid j = 1, \dots, N,\; \mathrm{Y}_j \in \mathbb{R}^{H \times W \times 3}\}$. For a visual condition $p$ (e.g., GZ2-annotated features; see Section~\ref{sec:data}), conditional realism $\mathcal{R}^p_C$ is quantified as below:  
\begin{equation}
    \mathcal{R}^p_C = \frac{1}{N}\sum_{j=1}^{N}\max_{i} \left(\mathcal{S}\left(f_\phi(\mathrm{X}_i), f_\phi(\mathrm{Y}_j)\right)\right), \quad i \in \{1, \dots, N'\},
\end{equation} 

where $f_\phi(\cdot): \mathbb{R}^{H \times W \times 3} \rightarrow \mathbb{R}^n$ is a SimCLR-based encoder~\cite[][]{CHEN2020} mapping images to an $n$-dimensional embedding space (implementation details in Appendix~\ref{appendix:simclr}), and $\mathcal{S}$ denotes cosine similarity. The global realism $\mathcal{R}$ averages $\mathcal{R}^p_C$ across all conditions $p \in \mathbf{P}$:  
\begin{equation}
    \mathcal{R} = \frac{1}{|\mathbf{P}|} \sum_{p\in\mathbf{P}} \mathcal{R}^p_C.
    \label{eq:average_realism}
\end{equation}

The conditional diversity $\mathcal{D}^p_C$ is defined as:  
\begin{equation}
    \mathcal{D}^p_C = \frac{1}{N \cdot N'} \sum_{i=1}^{N'} \sum_{j=1}^{N} \mathcal{S}\left( f_\phi(\mathrm{X}_i), f_\phi(\mathrm{Y}_j) \right),
    \label{eq:condition_diversity}
\end{equation}  
measuring average similarity between real and synthesized image pairs under condition $p$. The overall diversity $\mathcal{D}$ is computed analogously to realism:  
\begin{equation}
    \mathcal{D} = \frac{1}{|\mathbf{P}|} \sum_{p\in\mathbf{P}} \mathcal{D}^p_C.
    \label{eq:diversity}
\end{equation}  

It’s worth noting that the diversity metric measures how similar two images are when generated from the same morphology condition. As such, the smaller the diversity value, the greater the variety between the images.

To evaluate conditional consistency $\mathcal{C}^p$, we use the commonly adopted visual question answering (VQA) approaches \cite[e.g.,][]{YUSHIHU2023,CHOJAEMIN2023,ASTOLFI2024}. In detail, we train a series of classifiers $\mathbf{Q}_i$ tailored to different morphological features of galaxy images (hereafter, classifiers for VQA).

% \textbf{ -- (1) smoothness, (2) roundness, (3) edge-on disk orientation, (4) bulge shape, (5) bar presence, (6) spiral arm existence, (7) spiral arm tightness, (8) number of spiral arms, (9) bulge prominence, (10) presence of odd features, and (11) type of odd features (e.g., ring, merger, dust lane).}

These classifiers for VQA adopt a ResNet18 architecture and correspond to 11 key morphological questions derived from the Galaxy Zoo 2 (GZ2) question tree as Figure B3 shown, covering 11 distinct visual features, with the test set accuracies listed in parentheses -- (1) smoothness (80.8\%), (2) roundness (91.2\%), (3) edge-on disk orientation (94.5\%), (4) bulge shape (86.6\%), (5) bar presence (85.0\%), (6) spiral arm presence (89.8\%), (7) spiral arm tightness (69.0\%), (8) number of spiral arms (63.5\%), (9) bulge prominence (73.7\%), (10) presence of odd features (69.4\%), and (11) type of odd features (e.g., ring, merger, dust lane; 61.4\%). These accuracies reflect the performance of each classifier on a test set of real galaxies that were completely excluded from training, which is defined as the sum of true positives across all classes divided by the total number of samples.

Overall, these performance indices indicate that these classifiers for VQA achieve high accuracy for visually salient features (e.g., disk orientation, spiral arms) but relatively lower performance for more ambiguous or fine-grained categories (e.g., number or type of odd features), reflecting the intrinsic variability of these morphological traits. Also, the limited performance of classifiers for VQA maybe affect the evaluation. However, we have also measured average conditional consistency on real galaxy images as a benchmark, which is 0.895, shown in Figure~\ref{fig:step-3metrics}, which could enhance the reliability of consistency and provide a more robust basis for interpretation. 

This benchmark value in real images below 1 indicates that the classifiers are not perfect and may occasionally misclassify even real images. However, this does not diminish the utility of conditional consistency. The classifiers achieve accuracies ranging from approximately 60\% to 90\%, which ensures that consistency can provide quantitative measurements of feature correctness. By using the consistency of real images as a reference benchmark, we can meaningfully interpret and compare the consistency of generated images. The performance improvements observed in downstream classification tasks (see Section \ref{subsec:classical}) further validate this result.

% {\color{purple} ZC: May show the loss curve, key metric for those classifiers we used to evaluate the VQA? I think one question is for this part...} {\color{red}[done]}

Each classifier $\mathbf{Q}_i$ is designed to answer a specific morphological question as described above. They share the same ResNet18 architecture, and each classifier is trained from scratch independently, using standard cross-entropy loss and the Adam optimizer~\cite[][]{Kingma2015adam} with a learning rate of \(1\times10^{-3}\). Training proceeds until convergence with early stopping based on the validation loss. Each classifier is trained on a class-balanced subset of real galaxy images from GZ2, labeled according to the corresponding answers to its target question. For example, the classifier for identifying bulge shape is trained on a dataset comprising galaxies annotated as having a rounded bulge, a boxy bulge, or no bulge, with an equal number of samples per class. The dataset is randomly split into 80\% for training, 10\% for validation, and 10\% for testing.

% \textbf{Since the classifiers for VQA are trained to predict morphological features present in real galaxy images, we expect the conditional consistency score to approach 1 when these classifiers are applied to real images that clearly exhibit the desired features.} The conditional consistency $\mathcal{C}^p$ is then defined as the average accuracy of the classifiers $\mathbf{Q}_i$ under condition $p$:
\begin{equation}
    \mathcal{C}^p = \frac{1}{N}\sum_{j=1}^{N}\frac{1}{Q_j}\sum_{i=1}^{Q_j}\mathbb{I}\Big(\mathbf{Q}_i(\mathrm{Y}_j)=A_i\Big)
    \label{eq:condition_consistency}
\end{equation}
where $Q_j$ represents the number of classifiers employed for image $\mathrm{Y}_j$ ($11$ in this work), $A_i$ is the ground truth to classifier $\mathbf{Q}_i$ on image $\mathrm{Y}_j$, and $\mathbb{I}(\cdot)$ is the indicator function, which returns $1$ if the condition inside is true, and $0$ otherwise; it serves as a switch reflecting whether the stated condition holds. The global consistency $\mathcal{C}$ is the average of $\mathcal{C}^p$ across all conditions $p \in \mathbf{P}$:

\begin{equation}
    \mathcal{C} = \frac{1}{|\mathbf{P}|} \sum_{p\in\mathbf{P}} \mathcal{C}^p.
    \label{eq:consistency}
\end{equation}

Above metrics were tracked throughout the training iterations to monitor progressive improvements in image quality. As shown in Figure~\ref{fig:step-3metrics}, the realism of the generated images rapidly improve during the first 1,000 steps before continuing to grow steadily. Examples in Figure~\ref{fig:realvsgen} illustrates that synthetic images by GalaxySD closely resemble real galaxy images by capturing a wide range of morphological features—including spiral arms, dust lanes, and signs of mergers—according to the given conditions. Furthermore, Figure~\ref{fig:step-3metrics} reinforces that the diversity of the generated images increases continuously, demonstrating the model’s gradual mastery of detailed galaxy morphology, which is also evident in Figure~\ref{fig:realvsgen}.

\begin{figure}
    \centering
    \includegraphics[width=1\linewidth]{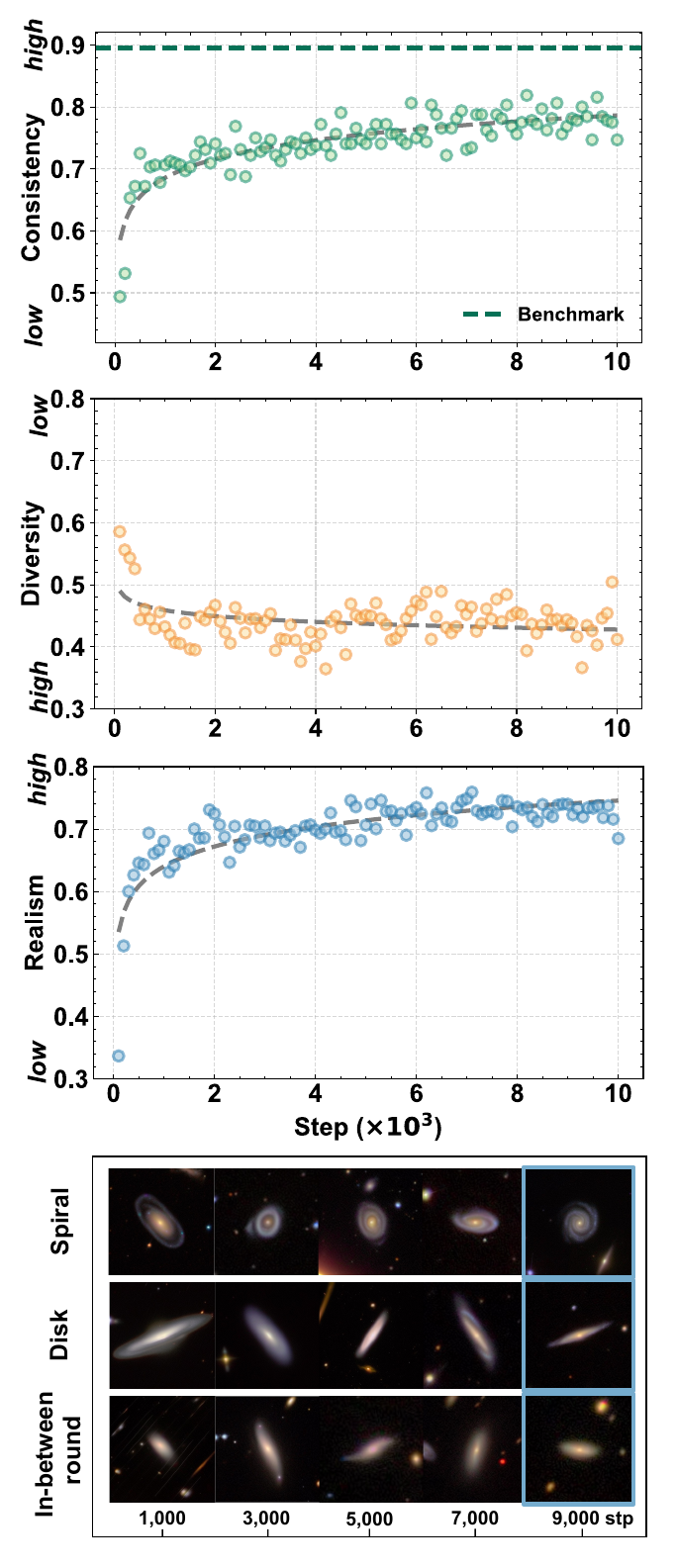}
    \caption{The quantified consistency, diversity and realism of generated galaxy images with training steps increasing, under average prompts. The benchmark line in the step vs. consistency figure is $y=0.895$, which denotes the consistency of real galaxy images under average prompts. The lower panel intuitively demonstrates that under various prompts, the more training steps evolve, the more real generated galaxy images are.}
    \label{fig:step-3metrics}
\end{figure}

\begin{figure}
    \centering
    \includegraphics[width=0.95\linewidth]{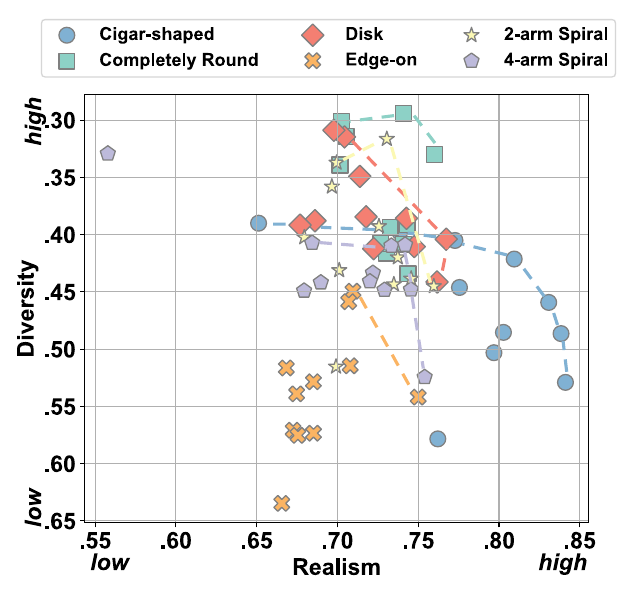}
    \caption{Pareto fronts showing the trade-off relationship between realism and diversity evaluation indicators of our conditional diffusion model. The dashed lines are Pareto fronts representing the frontiers of optimal solutions to the three evaluation metrics. Different colors represent different types of morphological prompts as top legend shows.}
    \label{fig:pareto-fronts}
\end{figure}

We also notice that there is a Pareto front between the realism and diversity of the generated images, as shown in Figure~\ref{fig:pareto-fronts}. This indicates that while GalaxySD can generate images with high realism, it may sacrifice some diversity in the process. This trade-off is common in generative models, where achieving high realism often leads to less diverse outputs. However, we find that our model maintains a good balance between these two aspects, producing images that are both realistic and diverse.

Overall, our fine-tuned conditional diffusion model, GalaxySD, successfully generates high-quality galaxy images that closely resemble real observations. The model exhibits strong realism, diversity, and consistency, as evidenced by the visual fidelity, morphological variety, and adherence to input conditions of the generated images. These results underscore the potential of our approach to enhance machine learning performance in various scientific applications, as detailed in the following sections.

\begin{table}[htbp!]
    \centering
    \begin{tabular}{llccc}
    \toprule
    \multicolumn{2}{c}{Tasks} & Purity & Completeness & F1-Score \\
    \midrule
    \midrule
    \multirow{3}{*}{\shortstack{Spiral\\(0.43)}} & Baseline & 0.67 & 0.45 & 0.54\\
    & Focal loss & 0.65 & 0.36 & 0.46\\
    & Ours & \textbf{0.69} & \textbf{0.70} & \textbf{0.70}\\
    \midrule
    \multirow{3}{*}{\shortstack{Disk\\(0.09)}} & Baseline & 0.81 & 0.52 & 0.63\\
    & Focal loss & 0.80 & 0.76 & 0.78\\
    & Ours & \textbf{0.83} & \textbf{0.77} & \textbf{0.80}\\
    \midrule
    \multirow{3}{*}{\shortstack{Bar\\(0.10)}} & Baseline & 0.59 & \textbf{0.95} & 0.73 \\
    & Focal loss & 0.61 & 0.85 & 0.71 \\
    & Ours & \textbf{0.67} & 0.86 & \textbf{0.75} \\
    \midrule
    \multirow{3}{*}{\shortstack{Bulge\\(0.07)}} & Baseline & 0.30 & 0.57 & 0.40 \\
    & Focal loss & 0.30 & 0.72 & 0.42 \\
    & Ours & \textbf{0.32} & \textbf{0.75} & \textbf{0.44}  \\
    \bottomrule
    \end{tabular}
    \caption{Performance comparison across classical binary classification tasks. The annotated fractions represent the proportion of positive samples in the full dataset. We focus on metrics for positive samples only, as negative sample metrics are uniformly high and do not effectively highlight differences between methods. Notably, machine learning models trained on our augmented dataset consistently outperform both baseline models and those using adjusted metrics.}
    \label{tab:classical-tasks}
\end{table}

\subsection{Improving Galaxy Morphology Classification with Synthetic Data}\label{subsec:classical}

As we have demonstrated that GalaxySD can generate high-fidelity galaxy images, we now show how these images can bring tangible scientific benefits. We first focus on classical binary classification tasks involving key galaxy morphologies, including spirals, edge-on disks, bars, and bulges (hereafter, classifiers for classical tasks). These features are critical for understanding various astrophysical processes, such as the role of spiral arms in star formation \cite[e.g.,][]{FOYLE2010,SCHINNERER2017,BINGQING2024}, the close relationship between the presence frequency of disk galaxies with $\Lambda$CDM cosmoloy \cite[e.g.,][]{NEELEMAN2020,NELSON2023,YANHAOJING2024,KOHANDEL2024}, the influence of bars on gas inflow and central starbursts \cite[e.g.,][]{FRIEDLI1994,FRASER2019,LIN2020,QIANHUI2023}, and the prominence of bulges in deciphering galaxy merger histories \cite[e.g.,][]{HOPKINS2010,FONTANOT2011,NEDKOVA2024}.

Despite these types of galaxies being relatively common in the Universe and having been widely studied over the past thirty years, their detailed physical scenarios remain a subject of debate. Accurately identifying them among billions of galaxies is of great scientific value. It not only helps us understand their formation mechanisms but also provides insights into cosmology \cite[e.g.,][]{HASLBAUER2022,WITTENBURG2023,HOPKINS2023}.

While these morphological features are much common compared to those rare-events such as strong lensing, tidal features, and dust lanes, they are still challenging to identify due to the lack of high-quality labeled datasets and the inherent class imbalance. As GalaxySD can produce a large number of realistic galaxy images with specific morphological features, we can use these synthetic images to enhance the training set for machine learning models. 

For each binary classifier for classical tasks, we fix the number of negative samples at 10,000 and sample positive samples proportionally to maintain the same class distributions as the GZ2 full dataset (containing 239,695 galaxies), with the detailed proportions listed in corresponding parentheses. Specifically, the training set for each classifier consists of a positive sample size of 4,300 for spiral (43\%), 900 for disk (9\%), 1,000 for bar (10\%), and 700 for bulge (7\%) features and 10,000 negative samples, respectively. These different positive sample ratios lead to varying training set difficulties across classifiers, with lower positive fractions (e.g., bulge) generally resulting in harder classification tasks. We here neglect the possible contaminations in the training set, as the contamination rate should not dominate the final results since we take strict and conservative data selection criteria as described in Section~\ref{sec:data}. 

Specifically, we augment the training set by generating a large number of synthetic galaxy images based on specific morphological prompts. The synthetic-augmented training set for the ``spiral" feature consists of 4,300 real galaxies (43\%) and 5,700 synthetic ones as augmented positive samples; for the ``disk", ``bar", and ``bulge" features, we use 900, 1,000, and 700 real images respectively, and augment them with synthetic images to reach 10,000 positive samples per task. In each case, 10,000 negative samples are randomly selected from real galaxies without the target feature, resulting in class-balanced training sets. Due to the limited number of real galaxy images from observation, the effective training set sizes of the synthetic-augmented and real datasets differ, according to the above description.

We evaluated the effectiveness of these augmentation-trained classifiers for classical tasks by comparing our model with two baselines: (1) a pure supervised learning classifier trained only on the GZ2 dataset and (2) a focal loss-trained classifier \cite[][]{lin2017focal} to address the class imbalance. Both methods are trained using only real images, with the same training set distribution as described in the previous paragraph, but without any synthetic augmentation.

% {\color{red} The focal loss is implemented by first computing the standard cross-entropy loss for each sample without reduction. Let \(p_t\) denote the predicted probability of the ground-truth class, obtained as \(p_t = \exp(-\mathrm{CE})\) from the cross-entropy loss \(\mathrm{CE}\). The final focal loss for each sample is then calculated as
% \[
% \mathrm{FL} = \alpha (1 - p_t)^\gamma \cdot \mathrm{CE},
% \]
% where $\alpha$ is a weighting factor to address class imbalance, and $\gamma$ is the focusing parameter that reduces the relative loss contribution from well-classified examples, thus emphasizing hard-classified examples. The per-sample focal losses are aggregated by taking the mean over the batch, which stabilizes training and ensures that the loss magnitude is independent of the batch size.} \CHENRUI{Is it necessary to write about how $\alpha$ and $\gamma$ are valued?}

The idea of the focal loss method is to modify the loss function, so that to address the issue of class imbalance and to improve the sensitivity of the model to hard-to-classify samples. Compared with the conventional cross-entropy loss, focal loss introduces a modulation mechanism that down-weights the contribution of easily classified examples while increasing the relative importance of misclassified ones. This property makes it particularly effective for tasks with skewed class distributions. The detailed formulation and implementation of focal loss are provided in the Appendix \ref{appendix:focal_loss}.

In these experiments, we split the prepared dataset for each task into a training set (90\%) and a test set (10\%) and further randomly select 10\% of the training data as a validation set. For both the real training set and the synthetic-augmented training set, we apply basic data augmentation operations, including rotation, cropping, flipping, zoom-in, and zoom-out, to enhance the diversity of the training samples and improve the model’s robustness to variations in galaxy orientations and scales. We then use the dataset to train a convolutional neural network (CNN) model with a ResNet18 backbone \cite[][]{HE2016} as the binary classifier for each of the four morphological features in baseline, focal loss and our synthetic-augmented methods. 

These classifiers for classical tasks are trained using a binary cross-entropy loss function except the focal loss case, and the Adam optimizer with a learning rate of \(1\times10^{-3}\). All hyperparameters are kept identical across the three methods to ensure a fair comparison. Rather than training for a fixed number of epochs or focusing on training loss, we apply the early stopping criterion based on the validation loss, terminating training once no further improvement is observed. Across the four synthetic-augmented classifiers for classical tasks, convergence was reached within roughly a dozen epochs, depending on the specific task and dataset. The classifier for spiral galaxies converged in 12 epochs, the classifier for disk-shaped galaxies in 13 epochs, the classifier for barred galaxies in 14 epochs, and the classifier for galaxies with bulges in 14 epochs.

We note that these binary classifiers for classical tasks differ from those used for consistency evaluation of our generation model, primarily in their training data. The former use synthetic images from our model for data augmentation and simulate the class distribution of the GZ2 dataset, thus are not class-balanced. In contrast, the latter are trained solely on real galaxy images with balanced positive and negative samples to reliably identify specific morphological features.

To evaluate the performance of these classifiers for classical tasks trained on our augmented dataset, we use three metrics: purity, completeness, and F1-score. Purity measures the proportion of correctly classified positive samples among all samples predicted as positive, while completeness measures the proportion of correctly classified positive samples among all actual positive samples. The F1-score is the harmonic mean of purity and completeness, providing a balanced measure of the classifiers' performance. 

%\TAO{Maybe we should provide the details of generative models based augmentation. E.g. What prompt are used? How many generated image are used?}

The results are presented in Table~\ref{tab:classical-tasks}. Our method outperforms both baselines across all four classification tasks, achieving the highest values for purity, completeness, and F1-score. For instance, in the spiral feature classification task, our method attains a purity of 0.69, completeness of 0.70, and F1-score of 0.70—surpassing the baseline model (0.67, 0.45, 0.54) and the focal loss model (0.65, 0.36, 0.46). Similar improvements are observed in the other three tasks. These findings demonstrate that our generative model effectively enhances the performance of machine learning models in identifying key morphological features within galaxy images, particularly for common galaxy morphology classification tasks.

The success of our method stems from the high quality and diversity of the generated images, which provide a rich source of synthetic data for training downstream ML models. Since image generation is a more complex task than binary classification, generative models are less prone to overfitting on systematic biases in the data. This enables them to improve the generalizability of downstream ML models. In astrophysics, data distributions are intrinsically imbalanced due to physical processes governing cosmic evolution. However, robust ML models require large training datasets to generalize well on real-data. Our method bridges this critical gap by generating synthetic data to address such imbalances.

The enhanced performance of our approach in real-world data classification tasks establishes a critical link between image quality evaluation—leveraging the heuristic metrics of realism, consistency and diversity outlined in Section~\ref{subsec:visual}—and scientific utility for research. This connection not only enables a more holistic assessment of synthesized images but also highlights the practical value of GalaxySD for advancing studies focused on galaxy morphology.

\subsection{Enhancing Early-type Dust Lane Galaxy Detection via Extrapolation on Rare Morphologies}\label{subsec:few-shot}

\begin{figure*}
    \centering
    \includegraphics[width=1\linewidth]{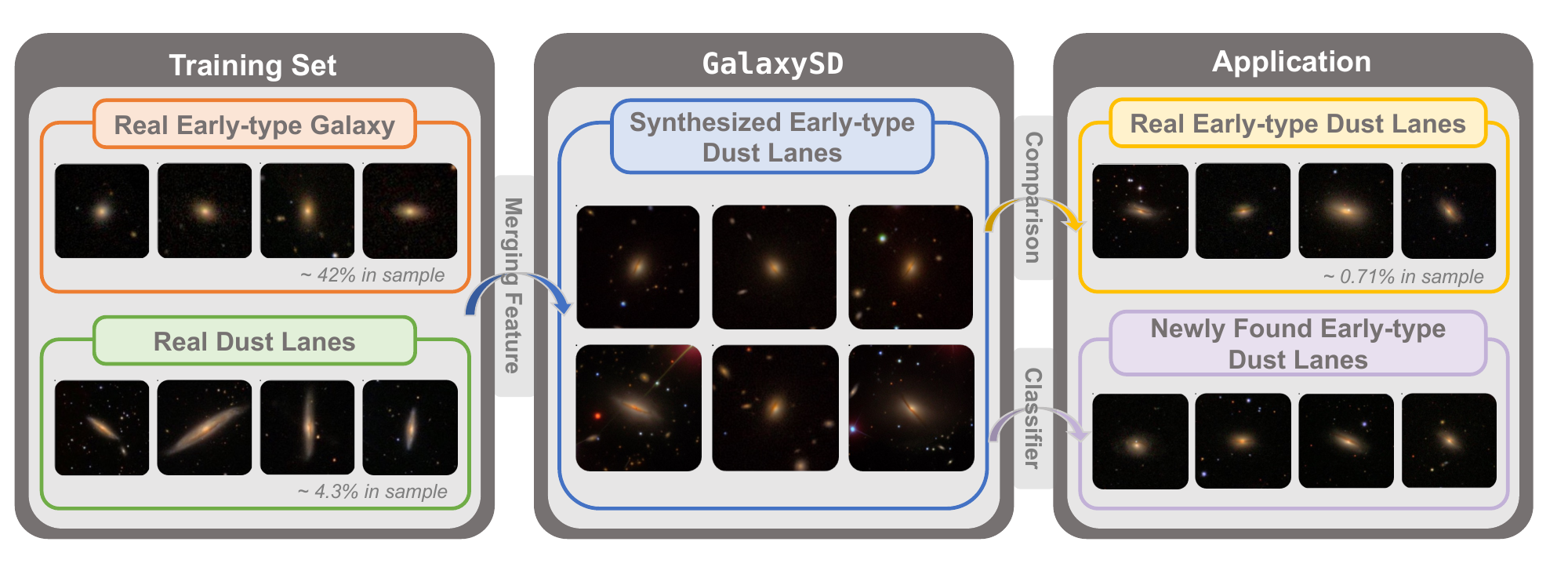}
    \caption{Generative extrapolation of early-type galaxies with prominent dust lanes (D-ETGs). While early-type galaxies are commonly quiescent, our generative model synthesizes D-ETGs by merging the morphology feature in early-type galaxy and star-forming feature (dust lane) in late-type galaxies. Both early-type galaxies and dust lanes are common in the universe, but early-type galaxies with dust lanes are rare since they are origin from recent merging events. Our generation model learns these two concepts from extensive datasets and extrapolates them to synthesize rare D-ETGs, which are excluded in training set. These synthesized D-ETG images are then used as training data for classifiers, facilitating the identification of 520 D-ETGs within a sample of 239,695 GZ2 images as shown in Figure~\ref{fig:summary_D-ETGs}.
    %Flowchart illustrating the image generation pipeline for early-type galaxies with dust lane (D-ETGs), which includes elliptical, lenticular and early spiral galaxies as shown. With abundant realistic D-ETGs generated by our model, the classifier trained on few real data and vast simulation data could perform well. As a result, the classifier work back on GZ SDSS and found a series of D-ETGs that were misclassified into other categories. The annotated proportions in each box refer to the number of those galaxies in the sample and also indicates how rare they are.
    }
    \label{fig:flowchart_earlytype_dustlane}
\end{figure*}

\begin{figure*}
    \centering
    \includegraphics[width=1\linewidth]{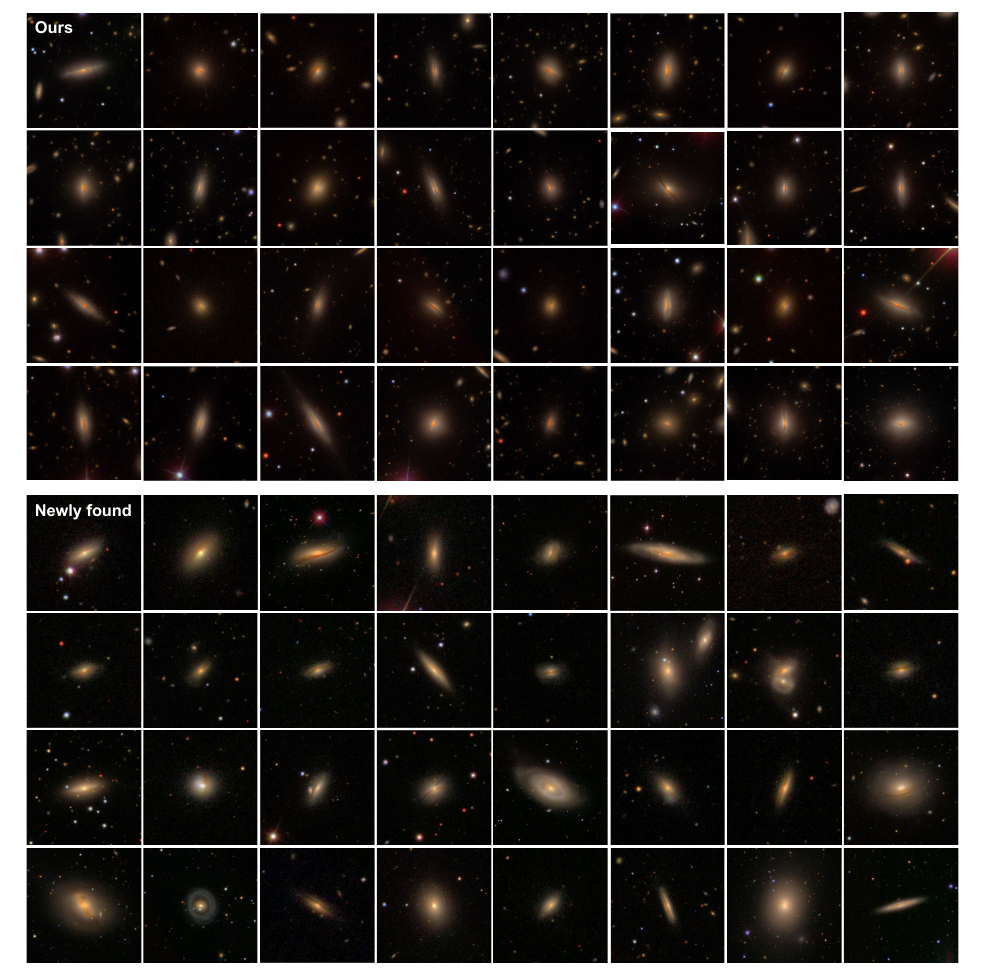}
    \caption{Examples of D-ETGs synthesized by our diffusion model (upper panel) and newly identified instances using machine learning models augmented with the synthesized data (lower panel). Our diffusion model generates high-fidelity D-ETG images by prompts like ``sdss, \{completely round:1.2\} galaxy, \{dust lane:1.3\}'' or ``sdss, in-between round galaxy, \{dust lane:1.2\}'', where the weights can vary reasonably within the range of 1.1 to 1.4. We manually assign weights to different morphology features through cross-attention mechanism to emphasize generation of desired features as in Section~\ref{sec:method}.}
    \label{fig:summary_D-ETGs}
\end{figure*}

\begin{table}
    \centering
    \begin{tabular}{lccc}
    \toprule
     & Purity & Completeness & F1-Score \\
    \midrule
    \midrule
    Baseline & 0.43 & 0.43 & 0.43 \\
    Focal loss &  0.48 & 0.16 & 0.24 \\
    Ours & \textbf{0.74} & \textbf{0.60} & \textbf{0.66} \\
    \bottomrule
    \end{tabular}
    \caption{Performance comparison on the D-ETG identification task. We follow the same experimental setting as in Table~\ref{tab:few-shot}. The synthesized D-ETG images can substantially enhance the downstream tasks as shown here.}
    \label{tab:few-shot}
\end{table}

As we have demonstrated that our generative model can generate galaxies with common morphologies, we now turn our attention to a more challenging task: identifying early-type galaxies with dust lanes (D-ETGs) by extrapolation on rare morphologies. The rarity of D-ETGs in the universe makes them difficult to be filtered out from large galaxy surveys. For instance, in the work of \cite{KAVIRAJ2012}, around 19,000 galaxies from GZ2 catalog was initially flagged by at least one user as having a dust lane. However, after visual inspection by two experts, only 352 galaxies were verified as D-ETGs.  Such low detection rate poses a challenge for machine learning models, as they often struggle to learn from such limited data. 

To address this issue, we leverage our generative model to synthesize additional D-ETG images, thereby augmenting the training set and improving the performance of machine learning models in identifying these rare objects. This task is difficult compared to the previous one, as the model needs to learn to extrapolate from common morphological features -- early-type features and dust lane features -- to extrapolate to the rare D-ETG morphology as shown in \autoref{fig:flowchart_earlytype_dustlane}.

We initiate our investigation with an experimental test. Using the D-ETG catalog from \cite{KAVIRAJ2012}, we aim to verify the effectiveness of our pipeline. In the experimental setup, we randomly select 100 galaxies from the D-ETG catalog in \cite{KAVIRAJ2012} and 10,000 galaxies from the GZ2 dataset to construct the training set, composing an experimental dataset with 1/100 postive / negative sample ratios, mimicking the data distribution in \cite{KAVIRAJ2012}. Similarly to the approach in Section~\ref{subsec:classical}, we use pure supervised learning without any additional adjustments as our baseline. 

Again, we introduce Focal Loss as an alternative baseline for dealing with imbalance classes as in Section~\ref{subsec:classical}. This method reweights the contributions of positive and negative samples in the loss function. We note that we defer detailed comparisons with more advanced few - shot algorithms and unsupervised classification algorithms to future studies, as their inclusion is beyond the scope of this work. More importantly, our methodologies are not in competition with these algorithms but can be used together to enhance these algorithms through data augmentation, a topic that we will discuss further in Section~\ref{sec:discussion}.

Despite the fact that the combination of early - type galaxy and dust lane features is rare in the universe, their individual components (early type galaxies, and galaxies with dust lane) can be easily found and are well - annotated in the GZ2 catalog. Our generative model learns these two dominant features and then extrapolates this concept to the context of early - type dust - lane galaxies. The synthesized D - ETGs, which are presented in Figure~\ref{fig:summary_D-ETGs}, mimic realistic D - ETGs and can be generated easily.  

As shown in \autoref{tab:few-shot}, injecting our synthesized images into the training pipeline improves the purity from 43\% to 74\% and the completeness from 43\% to 60\% at a positive sample ratio of 1/100. Focal Loss, while increasing the purity by 5\% by assigning more model weights to rare samples, leads to a decline in completeness, dropping from 43\% to 16\%. The result demonstrates the potential of our data augmentation based approach in enhancing the performance of galaxy morphology few-shot learning tasks.

\begin{figure}
    \centering
    \includegraphics[width=0.9\linewidth]{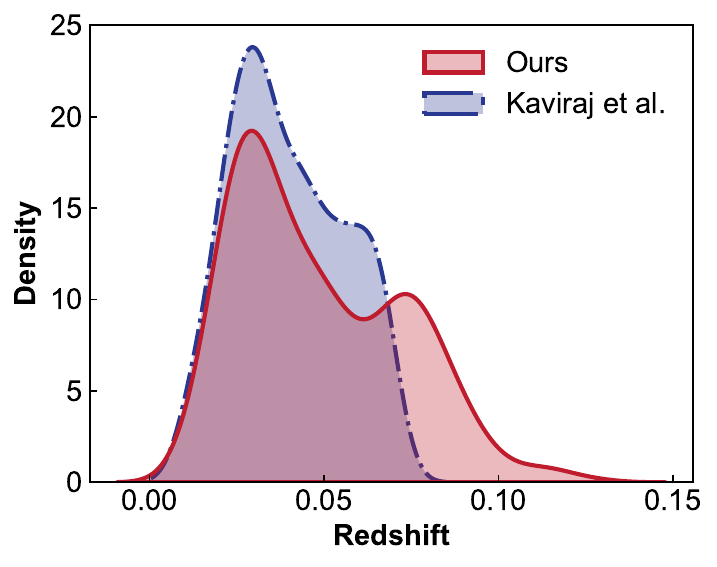}
    \caption{Kernel density estimates (KDE) of redshift distributions for D-ETGs identified by our method (red) and \cite{KAVIRAJ2012} (blue). Redshift values are obtained from the GSWLC \cite[][]{SALIM2016}. Our sample encompasses relatively higher redshifts, a regime that is more challenging for human volunteers to identify, thereby demonstrating the efficacy of our synthesized image-enhanced machine learning model in detecting rare galaxy populations.}
    \label{fig:D-ETG_redshift}
\end{figure}

\begin{figure*}[htbp]
\centering
\gridline{
\fig{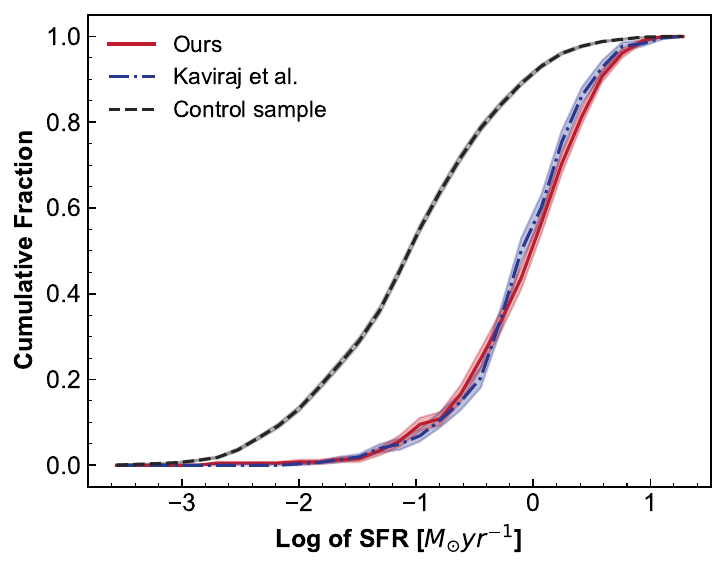}{0.48\textwidth}{(a) Star Formation Rate (SFR) Comparison}
\fig{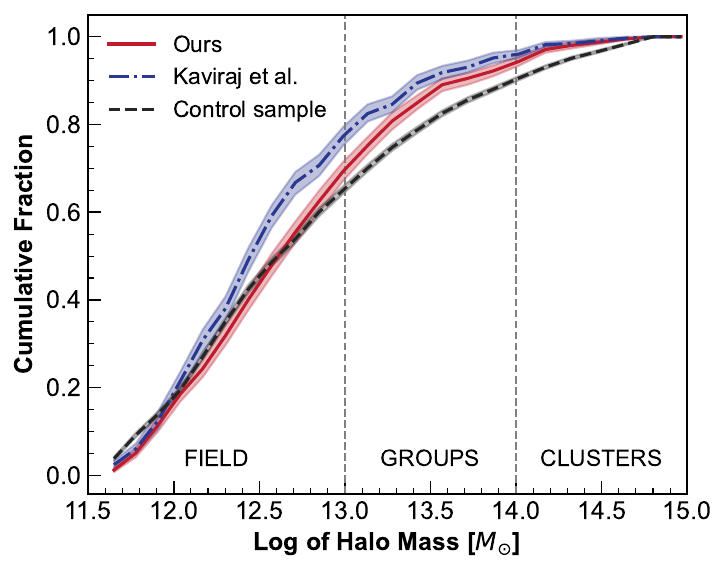}{0.48\textwidth}{(b) Halo Mass Comparison}}
\caption{Cumulative distributions of star formation rate (SFR; left) and halo mass (right) are shown for our D-ETG dataset, the \protect\cite{KAVIRAJ2012} sample, and a control ETG sample. SFRs are drawn from the GSWLC catalog \cite[][]{SALIM2016}, and halo masses are derived from the group catalog in \cite{XIAOHUYANG2012}; shaded regions denote 1$\sigma$ bootstrap errors for the cumulative distribution functions (CDFs). Despite their early-type morphology, our sample exhibits higher SFRs than control ETGs, aligning with \cite{KAVIRAJ2012} and indicating substantial star-forming activity. These galaxies preferentially inhabit field and group environments (consistent with \cite{KAVIRAJ2012}), suggesting a history of strong interactions. Together, these results imply their physical origin as minor merger remnants.}
\label{fig:sfr-halo-distributions}
\end{figure*}

Here, we apply our method to 239,695 Galaxy Zoo 2 (GZ2) images to identify previously undetected early-type dust lane galaxies (D-ETGs). We first construct training data: 10,000 synthetic D-ETGs as positive samples and 10,000 randomly selected normal galaxies as negative samples, to train a binary classifier for D-ETGs. Given the low occurrence of D-ETGs, we assume minimal contamination in the negative sample set. We follow the training procedure described in Section~\ref{subsec:classical}. After the initial training, the model was applied to the entire GZ2 dataset, identifying 8,184 positive candidates. These candidates were then visually inspected, and the classifier was retrained based on the validated results, yielding 2,488 refined positive samples. This iterative process was repeated once more, resulting in an additional 1,062 positive samples. All candidate samples in the three iteration training stages were visually validated by three authors (CM, ZS, and TJ). Final selection required agreement from at least two reviewers, resulting in a catalog of 520 high-confidence D-ETGs overlooked in \cite{KAVIRAJ2012}. We also note that a baseline model trained merely on the 352 D-ETGs from \cite{KAVIRAJ2012} yielded extremely low-purity results, rendering effective detection unfeasible.

These findings align with preliminary experiments in Table~\ref{tab:few-shot}, highlighting the critical role of synthetic images for enhancing rare object detection. The redshift distribution of our 520 newly identified D-ETGs is shown in Figure~\ref{fig:D-ETG_redshift}. Spectroscopic redshift measurements were extracted from the GSWLC catalog \cite[][]{SALIM2016}, originating from SDSS spectra \cite[][]{AHNCHRISTOPHER2014}. As shown, our samples exhibit a broader spread toward higher redshifts, making it challenging for humans to laboriously identify them among hundreds of millions of galaxies.

%\CHENRUI{But only 395 D-ETGs have physical parameter values such as redshift. How should it be written here？ Or the contributed catalog contains 467 D-ETGs with blank physical parameters.}

To ensure our sample shares similar physical properties as in \cite{KAVIRAJ2012}, we compare their Star Formation Rate (SFR) and halo mass in Figure~\ref{fig:sfr-halo-distributions}. The newly found ones were compared with those from \cite{KAVIRAJ2012} and a control sample of 4000 normal early-type galaxies (ETGs). The control sample was selected to match our sample in redshift and stellar mass distributions (see Appendix~\ref{appendix:sanity}). Stellar mass and SFR measurements are from GSWLC catalogs \cite[][]{SALIM2016}, derived via Spectral Energy Distribution (SED) fitting using CIGALE \cite[][]{CIGALE2009} with UV and optical photometry (\(0.15–0.9\,\mu\)m). The SED models assumed a two-component exponential star formation history, BC03 stellar populations \cite[][]{BC032003}, and Calzetti attenuation law\cite[][]{CALZETTI2000}. Halo masses were drawn from group catalogs in \cite{XIAOHUYANG2012}. We found that the D-ETGs in this work exhibit higher star formation rates than their normal counterparts, consistent with the samples in \cite{KAVIRAJ2012}. These galaxies are also more likely to reside in field and group environments, suggesting their origin from merger events \cite[][]{PEARSON2024}.

Our approach ultimately identifies 520 D-ETGs that were overlooked in \cite{KAVIRAJ2012}, thereby opening up new avenues for understanding merger events and dust properties in nearby galaxies. We provide this new sample as an open-access catalog to facilitate further studies of early-type dust lane galaxies, as described in Section \ref{sec:data_avail}. The catalog contains, for each of the 520 D-ETGs, the SDSS DR7 object identifier, the corresponding filename as an asset identifier (asset\_id), RA, DEC, and subsample identification.

\section{Discussion}\label{sec:discussion}

Our work demonstrates the potential of diffusion models to generate high-fidelity galaxy images that closely mimic real astronomical observations while enforcing specified morphological characteristics. By integrating these synthetic images into machine learning (ML) training pipelines, we enhance the performance of diverse scientific tasks—from classical morphological classifications (e.g., bars, spirals, disks, bulges) to the more challenging detection of rare objects such as dusty early-type galaxies (D-ETGs). While state-of-the-art hydrodynamical simulations remain computationally intensive and often struggle to resolve fine-scale observational details, our diffusion models can synthesize large datasets of observationally realistic images on a single GPU. These synthetic galaxies facilitate the identification of analogous systems in real astronomical surveys, forming a real-to-synthetic-to-real workflow that provides new observational constraints on those elusive phenomena. 

Our methodology is readily adaptable to other galaxy populations of high scientific relevance, such as strong lensing systems \cite[e.g.,][]{BROWNSTEIN2012,ORIORDAN2025,DENGLIMENG2025}, red spirals \cite[e.g.,][]{MASTERS2010,FUDAMOTO2022,JIANTONGCUI2024}, tidal features \cite[e.g.,][]{HOOD2018,KHALID2024,RUTHERFORD2024}, ultra-diffuse galaxies \cite[e.g.,][]{JIAXUANLI2023,LEYAO2025,BUZZOUDG2025} and etc. Detailed implementation for each system lies beyond the scope of this study and will be explored in future investigations. Additionally, for modern large-scale sky surveys, substantial quantities of high-fidelity synthetic data encompassing diverse galaxy types and observational conditions are critical for developing and validating data processing pipelines. Our diffusion model, therefore, can efficiently generate such validation data, thereby contributing to this domain as shown in previous studies \cite[e.g.,][]{GALSIM2015,LANUSSE2021}. By further integrating instrumental characteristics (e.g., noise level, PSF) and physical properties (e.g., stellar mass, dynamical information), we envision a unified “simulator” that captures complex correlations within multi-dimensional observational datasets. 
%This framework paths the way for generative models to reproduce and extrapolate existing observational data to address diverse scientific objectives.

The implications of our work extend to a broad spectrum of astrophysical research domains. For example: (1) In radio astronomy, projects such as Radio Galaxy Zoo \cite[e.g.,][]{BANFIELD2015,TANG2020,BOWLES2023} could leverage our pretrained models to classify intricate radio galaxy morphologies, such as bent jets or diffuse lobes. (2) For near-Earth object (NEO) detection, our framework could generate synthetic NEOs with realistic trajectories and observational systematics to enhance detection algorithms for hazardous asteroids \cite[e.g.,][]{DUEVDMITRY2019,IRURETAGOYENA2025}. (3) In planetary geology, synthetic surface maps of icy moons could train terrain ML models for upcoming missions \cite[e.g.,][]{BHASKARA2024}. The same workflow can also be extended to other data modalities, such as spectra, integral field unit (IFU) data, and others.
%These examples underscore the versatility of our generative approach in bridging observational gaps across scales—from galactic structures to planetary surfaces. 

Compared to recent efforts to build foundation models in astrophysics (e.g., AstroPT \cite[][]{ASTROPT2024,ASTROPT2025}, AstroCLIP \cite[][]{ASTROCLIP2024}), our work adopts a complementary strategy to address data scarcity. While such pretrained models encode compact representations of observed galaxy images through self-supervised learning, their deployment to specific scientific tasks often requires extensive fine-tuning with domain-specific labels. 

Here, our framework offers dual synergies: first, by generating high-quality labeled datasets to streamline fine-tuning for targeted applications, and second, by augmenting pretraining itself through synthetic samples that populate underrepresented regions of high-dimensional pixel space (e.g., rare morphologies, extreme physical parameters and instrument systematics). This symbiotic relationship between generative and foundational models promises to accelerate discoveries in data-driven astrophysics.

Despite our diffusion model demonstrating capabilities akin to a world simulator—for example, generating realistic dusty early-type galaxies (D-ETGs) by combining morphological features from early-type galaxies and disky galaxies with dust lanes—this data-synthetic methodology remains confined to the framework of association modeling. As such, a substantial gap persists between these diffusion models, a challenge that extends even to state-of-the-art video generation models like Sora\footnote{\hyperlink{https://openai.com/index/sora/}{https://openai.com/index/sora/}}. By relying solely on associative relationships between observed features, such generative models cannot address counterfactual ``what if" questions—for instance, ``What would this galaxy look like if traced back to its formation epoch?" or ``How would its morphology differ under altered cosmological conditions?" \cite[][]{PEARL2009,PERAL2010}.

As such, we explicitly frame our diffusion model as a surrogate for data augmentation under known feature associations, rather than a tool for physical simulation or causal inference. Its utility lies in expanding the diversity of observable feature combinations within the constraints of existing astronomical knowledge. For questions requiring causal modeling of cosmic processes, complementary physical simulation frameworks—incorporating cosmological evolution and subgrid physics will remain indispensable \cite[e.g.,][]{CRAIN2015,NELSON2019}.

Looking ahead, generative models like ours are poised to redefine the frontiers of scientific discovery by enabling imaginations at cosmic scales. By synthesizing physically plausible yet observationally rare phenomena—from extreme galaxy mergers \cite[e.g.,][]{ATHANASSOULA2019} to hypothetical Pop III galaxies \cite[e.g.,][]{TRUSSLER2023,XIAOJINGLIN2023} -- these models can serve as hypothesis generators, guiding telescopes toward uncharted regions of parameter space. Furthurmore, by blending data-driven synthesis with physical priors, we envision a future where generative models act as collaborative partners in the scientific process—not just filling data gaps, but actively expanding the scope of questions we can ask about the universe.

\section{Conclusions}\label{sec:conclusion}

In this study, we propose employing a conditional diffusion model as a surrogate for synthesizing galaxy images. Our research demonstrates that this diffusion model can generate high-fidelity galaxy images, closely adhering to specified morphological conditions while maintaining low computational cost. These synthesized images effectively address the data scarcity issue in astrophysical machine learning training pipelines. As evidenced by our experiments on tasks such as galaxy morphology classification and rare object detection, incorporating these augmented training data into machine learning training pipelines yields tangible scientific benefits. We summarize our contributions as follows:
\begin{enumerate}
\item We fine-tuned the pretrained conditional diffusion model \texttt{Stable-Diffusion-v1-5} using high-quality morphology label - galaxy image pairs from the Galaxy Zoo 2 dataset. The fine-tuned model demonstrates excellent performance in synthesizing high-fidelity images that closely follow morphological instructions.
\item We quantitatively evaluated the quality of the generated images from three distinct perspectives: (1) realism (visual fidelity relative to real observations), (2) diversity (morphological variety across outputs), and (3) consistency (adherence to input conditions), thus providing a novel evaluation method for astronomical image simulation. Our results show that after sufficient training, the diffusion model can generate realistic galaxy images that are consistent with input morphology conditions and exhibit diversity. We also explored the trade-off between realism and diversity, which can serve as a valuable reference for future applications.
\item We demonstrated that integrating the synthesized images into machine learning pipelines can improve the purity and completeness of machine learning model performance by up to 30\% in classical morphology classification tasks, such as spiral/disk/bar/bulge classification, compared to the baseline. This highlights the model's potential in overcoming data shortages in machine learning training pipelines.
\item We further demonstrated that our diffusion model functions as a “world simulator” surrogate, enabling generative extrapolation from well-observed galaxies to rare or even unobserved domains. Using early-type galaxies with prominent dust lanes as a test case (accounting for approximately 0.1\% of the Galaxy Zoo 2 dataset and excluded from the training data beforehand), we showed that our generative model can successfully synthesize these galaxies and enhance downstream machine learning models. Our method led to 520 additional dusty early-type galaxies from the same dataset, doubling the sample size compared to previous studies, which only reported 352 samples. This newly created catalog holds scientific value as a proxy for studying merger remnants and dust properties.
\end{enumerate}

While state-of-the-art hydrodynamic simulations seek to reproduce the Universe’s formation via first-principle calculations, our focus here is to empirically model it by harnessing the exponential growth of astronomical observational data. By employing conditional diffusion models to generate large quantities of realistic galaxy images, we supply rich, diverse training samples for machine learning (ML) algorithms and bridge gaps in observational datasets.

This framework not only improves the performance of ML models in tasks like galaxy morphology classification and rare object detection but, more crucially, enables scientists to explore physical processes not yet fully understood through generative extrapolation. In doing so, it constructs a novel bridge between theoretical frameworks and observational evidence, accelerating our comprehension of the Universe’s evolution and structural formation through a purely data-driven lens.

\section{Data Availability}\label{sec:data_avail}

The data used in this study, including the preprocessed Galaxy Zoo 2 (GZ2) dataset\footnote{\href{https://zenodo.org/records/15669465}{Training dataset: https://zenodo.org/records/15669465}} and the early-type dust lane galaxy catalog we contributed\footnote{\href{https://zenodo.org/records/15636756}{D-ETG catalog: https://zenodo.org/records/15636756}}, are publicly available now. The GZ2 dataset can be accessed at \url{https://data.galaxyzoo.org/}, which includes morphological classifications of galaxies from the Sloan Digital Sky Survey (SDSS). The early-type dust lane galaxy catalog for comparison is available in \cite{KAVIRAJ2012}. 

% The synthetic galaxy images generation in this study, along with the trained diffusion model and evaluation scripts, are available at \url{https://github.com/chenruiRae/GalaxySD}. This repository also includes the preprocessing scripts used to prepare the GZ2 dataset and the prompt structure for conditioning the diffusion model.

The model pipeline in this study is available at the GalaxySD\footnote{\mbox{\href{https://github.com/chenruiRae/GalaxySD}{https://github.com/chenruiRae/GalaxySD} with \href{https://zenodo.org/records/17112014}{DOI: 10.5281/zenodo.17112014}}} Github repository. And the trained diffusion model weights\footnote{\mbox{\href{https://huggingface.co/CosmosDream/GalaxySD}{https://huggingface.co/CosmosDream/GalaxySD} with \href{https://huggingface.co/CosmosDream/GalaxySD?doi=true}{DOI: 10.57967/hf/6479}}} are available as well. For any additional information or specific requests regarding the data or code, please feel free to contact us.

\section{Author Contributions}

% \ZECHANG{I prefer a author contribution section here since this work involved contributions from three students,  despite astronomy research community didn't encourage such collaborations (only one first-author and one corresponding author are recognized since we are not targeted at some high-profile journals. I perfer designing a high-profile project in next paper, since the review process for those journals are quite annoying and we need to rewrite the paper if rejected, which will introduce a lot of works). To recognize contributions from ZS and TJ, I think we would better clarify author contributions here.}

CM conducted experimental runs, created scientific visualizations, and contributed to draft revision and writing. ZS oversaw program design, developed code for the embedding model, and led manuscript writing and revision efforts. TJ gave the initial implementation of the diffusion model fine-tuning, participated in scientific discussions, and provided critical suggestions for manuscript refinement. ZC, YST, and SH offered expert scientific and technical guidance, along with constructive feedback on the manuscript. YST additionally contributed to the selection of the D-ETG research topic and the D-ETG catalog. MYL developed the visual inspection program utilized in the analysis.

\section{Acknowledgments}

ZS and SH acknowledge support from the National Natural Science Foundation of China (Grants No. 12273015 \& 12433003) and the China Crewed Space Program through its Space Application System. ML and ZC acknowledge support from the National Key R\&D Program of China (grant no. 2023YFA1605600) and Tsinghua University Initiative Scientific Research Program (No. 20223080023). CM acknowledge support by the Major Key Project of Peng Cheng Laboratory. The authors thank Hongming Tang, Yanhan Guo and Huiling Liu for their valuable discussions and constructive comments. We also appreciate the helpful technical feedback provided by \hyperlink{https://www.morphstudio.com}{Morph AI Inc}. The authors acknowledge the Tsinghua Astrophysics High-Performance Computing platform at Tsinghua University for providing computational and data storage resources that have contributed to the research results reported within this paper.

\appendix
\renewcommand\thefigure{\Alph{section}\arabic{figure}} % A1
\setcounter{figure}{0}

\section{SimCLR as a embedding tool for galaxy image}\label{appendix:simclr}

As our evaluation metric described in Section~\ref{sec:result} involves an encoder $f_{\phi}:\mathbb{R}^{\mathrm{H}\times\mathrm{W}\times 3}\rightarrow\mathbb{R}^{\mathrm{L}}$ to map high-dimensional galaxy images into low-dimensional latent representations, we here detail the representation learning technique used in this work -- SimCLR here.

SimCLR \cite[][]{CHEN2020} is a widely-used self-supervised learning technique in computer vision. The core idea behind, named as contrastive learning, is to construct a series positive and negative pair through data augmentation, and then train a neural network to pull together representations of two differently augmented views from the same galaxy image, while simultaneously pushing apart representations from distinct images. This contrastive mechanism drives the encoder to learn invariant features that robustly capture the morphological details of galaxies, thereby creating a highly discriminative embedding space. As a result, the learned representations not only improve the performance in downstream tasks such as image synthesis and classification but also enhance the overall ability to discern subtle astrophysical features. Compared to other representation learning techniques such as masked image modeling \cite[e.g.,][]{ZHENDAXIE2021} and variational autoencoders \cite[e.g.,][]{Kingma2013}, contrastive learning balances both model performance and the cost to training the model and already has wide applications in astrophysics \cite[e.g.,][]{STEIN2022,DESMONS2024,MOHALE2024}. 

Formally, let $\mathbf{x}_i$ be an image of a galaxy, and let $\mathbf{x}_i^a$ and $\mathbf{x}_i^b$ be two augmented views of $\mathbf{x}_i$. These views form a positive pair $(\mathbf{x}_i^a, \mathbf{x}_i^b)$. For a batch of $N$ images, we generate $2N$ augmented views, resulting in $N$ positive pairs and $2N(2N-2)$ negative pairs. The goal is to learn an encoder $f_\phi$ that maps images to a latent space where positive pairs are close and negative pairs are far apart.

The contrastive loss, often referred to as the InfoNCE loss, is defined as:
\[
\mathcal{L}_{\text{contrastive}} = -\sum_{i=1}^{N} \log \frac{\exp(\text{sim}(f_\phi(\mathbf{x}_i^a), f_\phi(\mathbf{x}_i^b))/\tau)}{\sum_{j=1}^{2N} \mathbb{I}_{[j \neq i]} \exp(\text{sim}(f_\phi(\mathbf{x}_i^a), f_\phi(\mathbf{x}_j))/\tau)},
\]
where $\text{sim}(\cdot, \cdot)$ denotes the cosine similarity between two vectors, $\tau$ is a temperature parameter, and $\mathbb{1}_{[j \neq i]}$ is an indicator function that is 1 if $j \neq i$ and 0 otherwise.

By minimizing this loss, the encoder learns to produce representations that are invariant to the augmentations applied to the same galaxy image while being discriminative enough to distinguish between different galaxies. This learned representation can then be used for various downstream tasks, such as galaxy morphology classification, anomaly detection, and image synthesis, providing a robust foundation for further analysis.

To implement the SimCLR framework, we utilize a ResNet18 architecture as our encoder $f_\phi$. ResNet18, a widely-used convolutional neural network, is chosen for its balance between depth and computational efficiency, making it suitable for extracting rich features from galaxy images. The network is initialized with weights pretrained on ImageNet to leverage transfer learning benefits. We train the model on the Galaxy Zoo dataset, which includes 300,000 annotated galaxy images from the Sloan Digital Sky Survey (SDSS). Each image undergoes a series of augmentations, such as random cropping, flipping, and color jittering, to generate positive pairs. The model is trained for 200 epochs with a batch size of 256 and a learning rate of 0.001, using the Adam optimizer. The temperature parameter $\tau$ is set to 0.5. 
\begin{figure}
    \centering
    \includegraphics[width=0.54\linewidth]{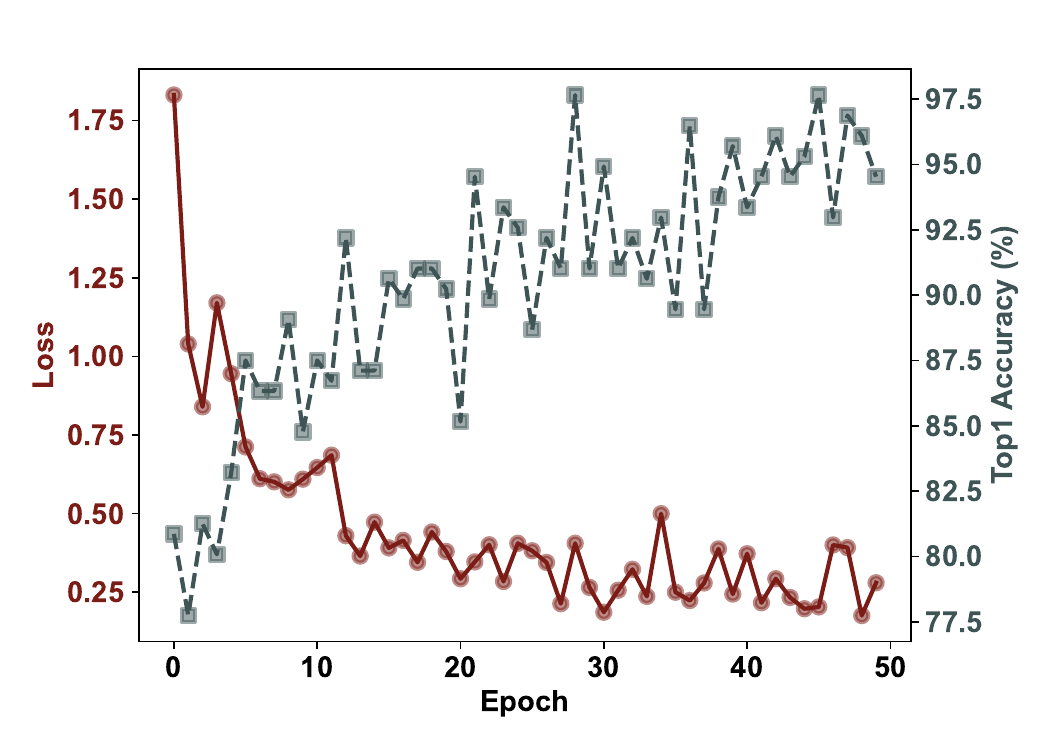}
    \caption{The contrastive loss and top-1 accuracy for evaluation along epochs when training SimCLR. Our representation learning model converges stably during the training progress.}
    \label{fig:loss_simclr}
\end{figure}

\begin{figure}
    \centering
    \includegraphics[width=0.45\linewidth]{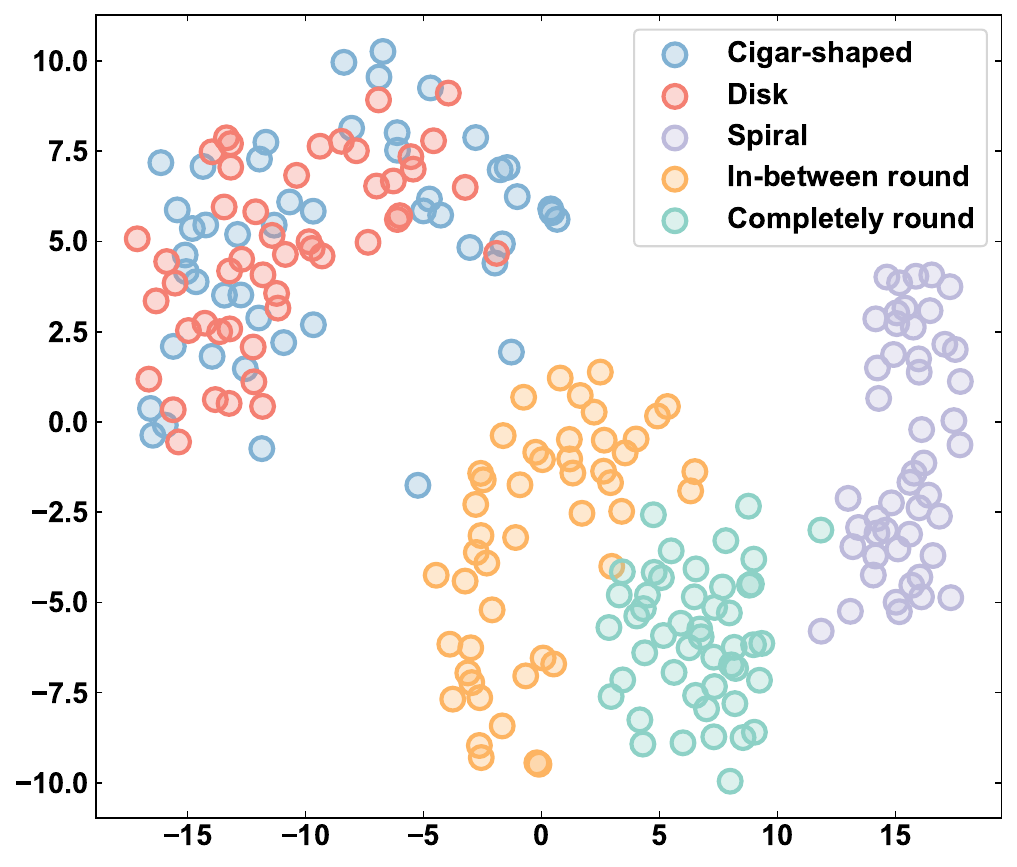}
    \caption{t-SNE visualization of the learned embeddings from the SimCLR model. The points are colored by different morphological features as shown in Figure~\ref{fig:realvsgen}. Different morphologies are clearly distinguished within the latent space, demonstrate the effectiveness of our representation learning model.}
    \label{fig:t_sne}
\end{figure}

The training process effectively minimizes the contrastive loss, as evidenced by the steady decrease in loss values and the increase in cosine similarity between positive pairs over epochs, as shown in Figure~\ref{fig:loss_simclr}. This indicates that the encoder is successfully learning to produce invariant representations for augmented views of the same galaxy image while distinguishing between different galaxies. The t-SNE visualization in Figure~\ref{fig:t_sne} further demonstrates that the learned embeddings form distinct clusters corresponding to different galaxy morphologies, validating the effectiveness of the SimCLR framework in capturing meaningful features. The aim of this training is to create a robust embedding space that can be used to evaluate the quality of generated galaxy images, ensuring that they closely resemble real observations and adhere to the specified morphological conditions.

\section{Heriarichcal Morphological Labelling System in GZ2}\label{appendix:labellingsystem}

The GZ2 labeling system uses a hierarchical framework starting with broad early/late-type galaxy distinctions, followed by detailed features like bars, dust lanes, and merger signatures. This structured approach enables systematic annotation of both global and fine-grained morphological attributes, as shown in Figure~\ref{fig:questiontree}.

\begin{figure}[htbp!]
    \centering
    \includegraphics[width=1\linewidth]{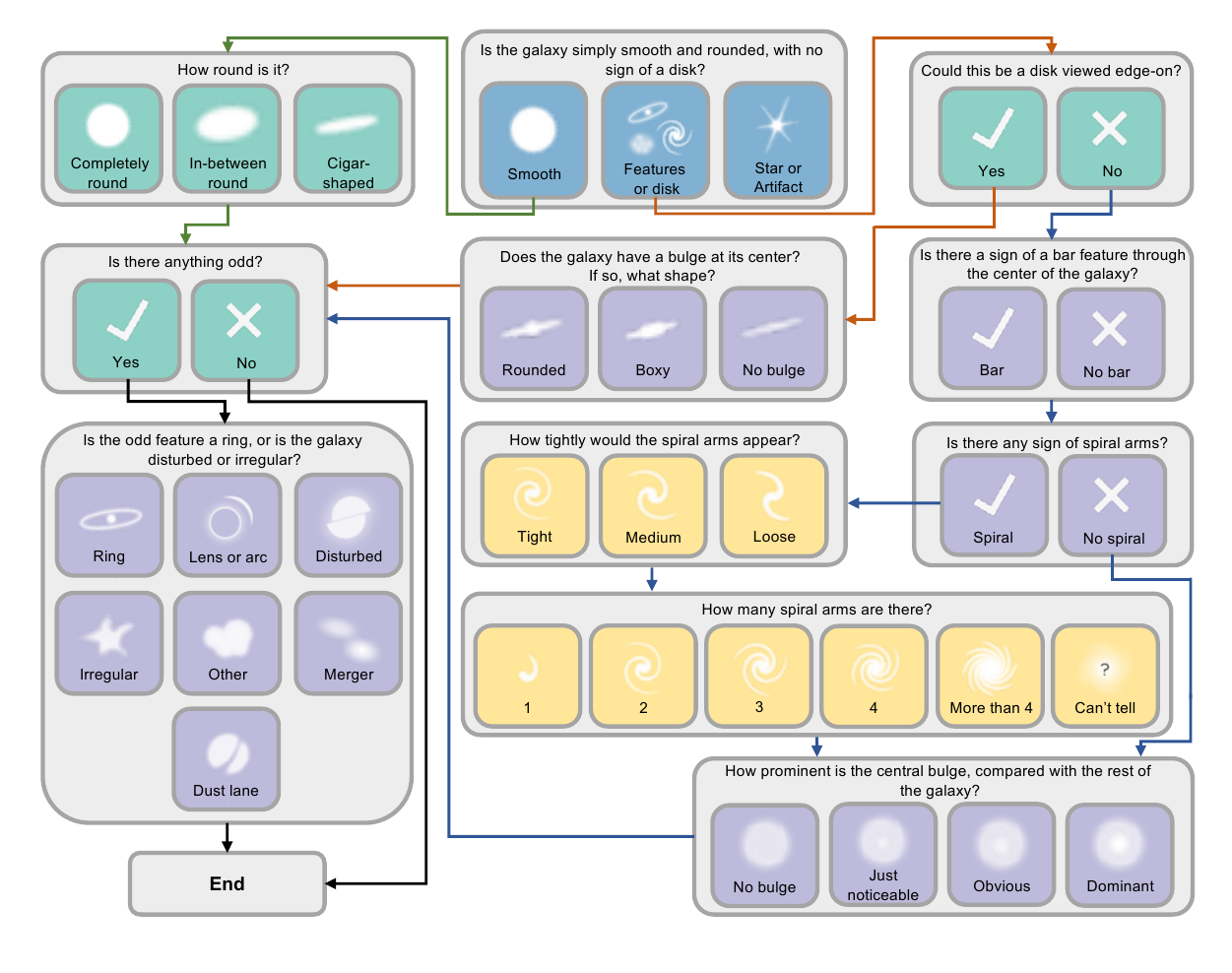}
    \caption{The question tree we use to annotate galaxy morphologies was initially proposed by \citet{Willett2013gz2sdss}. This hierarchical framework includes both high-level early/late-type classifications and detailed morphological features (e.g., dust, or mergers).}
    \label{fig:questiontree}
\end{figure}

\section{Weighted Prompt of dust lane feature}\label{appendix:weighted_prompt} 

As we have described in Section~\ref{subsec:cross_attention}, the weights are incorporated into the conditional text embeddings to guide the generation process. This strategy enables us to systematically enhance a target feature (e.g., dust lane) while simultaneously monitoring its impact on other morphological attributes. To quantitatively evaluate and determine the appropriate weights for emphasizing specific features during generation, we conducted controlled experiments in which the weight assigned to the ``dust lane'' feature was gradually varied. By analyzing how the generated images respond to different weight values, we first show how the weight quantitatively influences the desired feature, as well as how increasing or decreasing this weight affects other correlated features. This analysis provides a principled basis for how to assign reasonable weight values in prompts when emphasizing particular morphological characteristics.

\begin{figure}
\centering
\gridline{
\fig{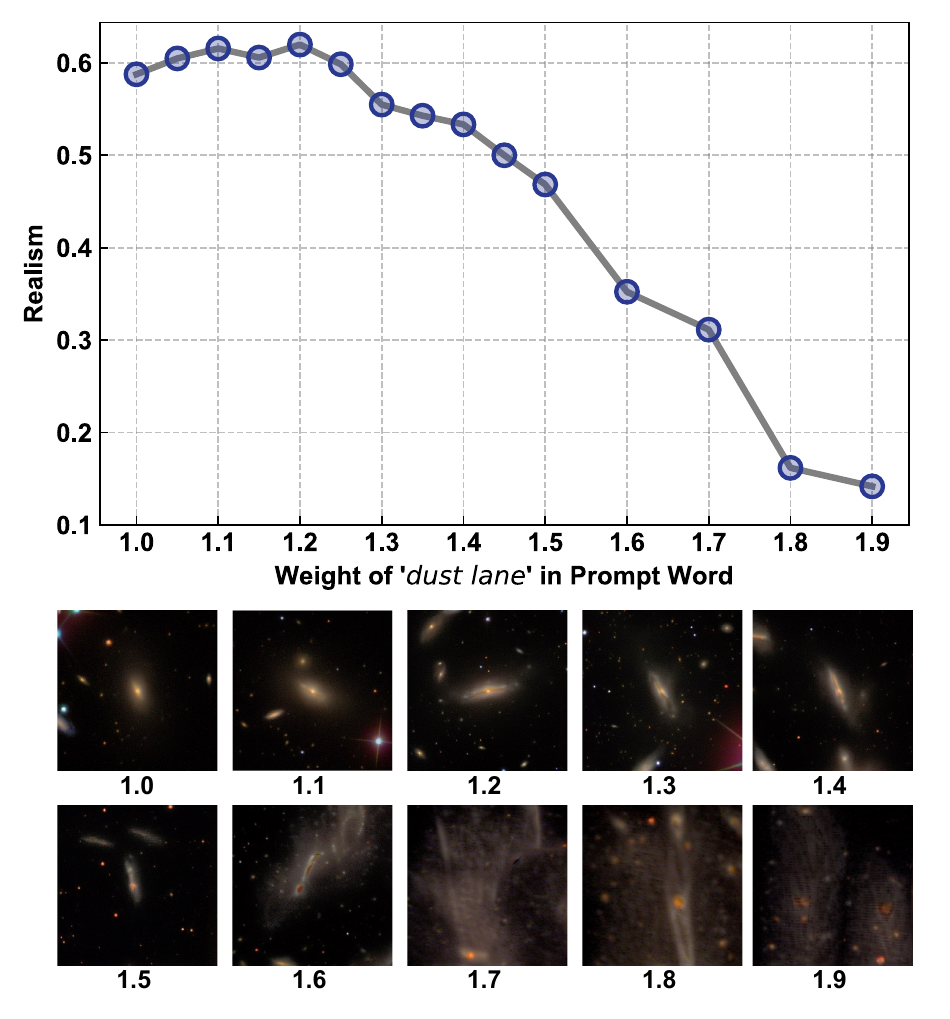}{0.48\textwidth}{(a) ``in-between round galaxy, \{dust lane:X\}''}
\fig{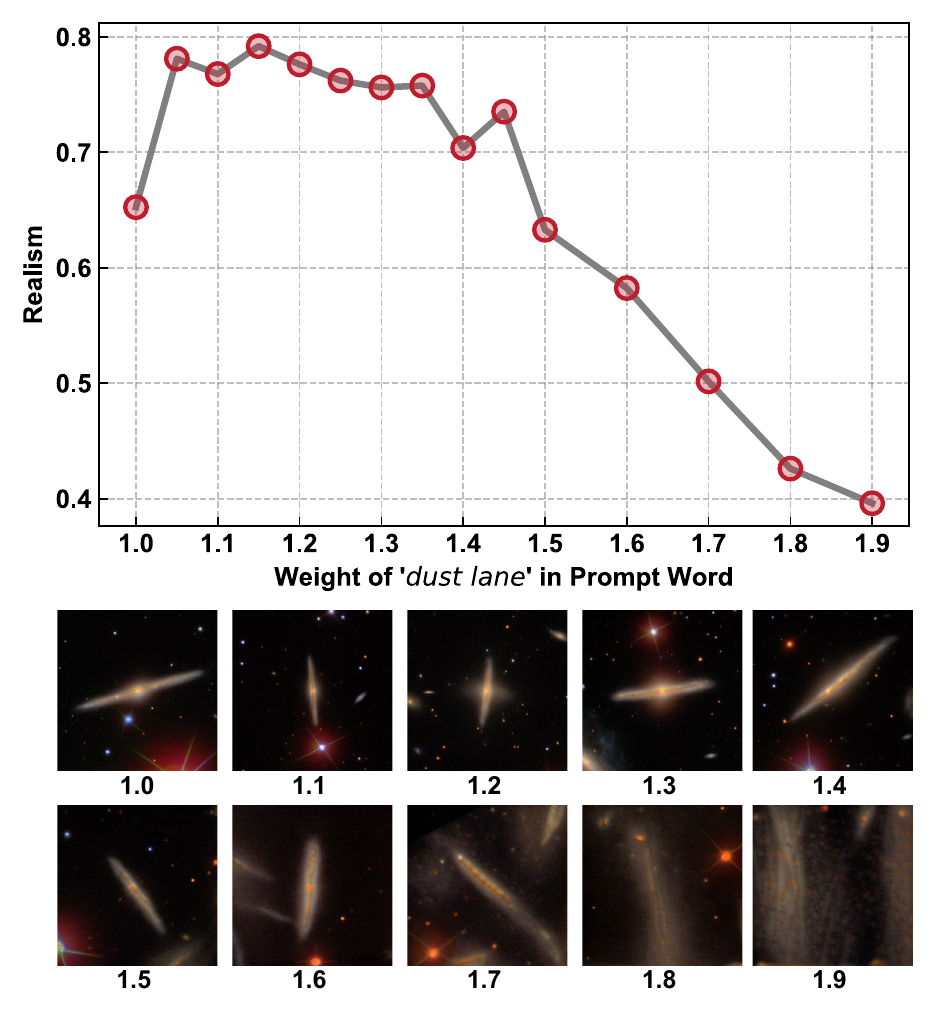}{0.48\textwidth}{(b) ``edge-on galaxy, \{dust lane:X\}, with rounded edge-on bulge''}
}
\caption{Effect of varying the weight of the “dust lane” feature in the text prompt on generated galaxy image realism scores. The (a) and (b) are two different prompts. In each sub-figure, the top panel shows the tendency of realism scores with respect to the “dust lane” weight (from 1.0 to 1.9) in the prompt word, peaking near 1.2 and then gradually declining as the weight increases from 1.5 to 1.9. The bottom two rows of images are example generated galaxies corresponding to each weight value annotated below. Increasing the weight initially strengthens the presence of dust lanes, but excessively high weights ($\geq$ 1.6) introduce unrealistic textures and artifacts, reducing the perceived realism.}
\label{fig:realism_for_prompt_weight_a}
\end{figure}

\begin{figure}
    \centering
    \includegraphics[width=1\linewidth]{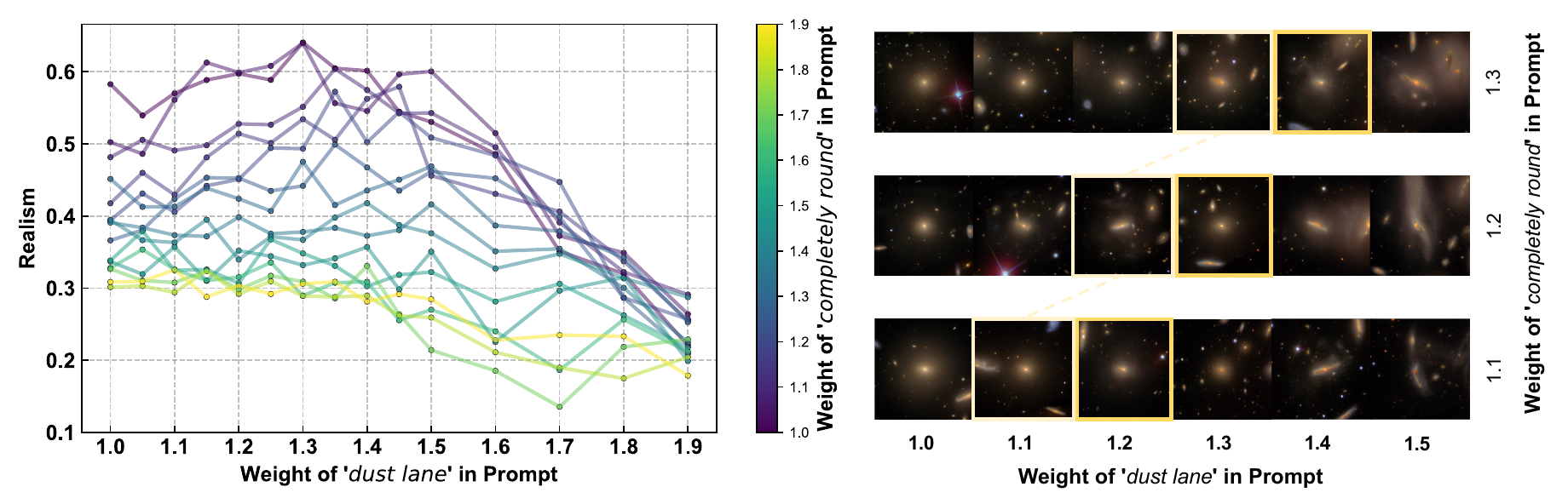}
    \caption{Effect of varying the weight of the “dust lane” and “completely round" feature in the text prompt ``\{completely round:X\} galaxy, \{dust lane:Y\}" on generated galaxy image realism scores, where X and Y are weights to the corresponding feature. The integration of these two features contributes to the final D-ETG generation. The highest realism scores of generated D-ETG images occur in the region where X $\in [1.1,\,1.4]$ and Y $\in [1.05,\,1.3]$. In the right panel, the boundary is formed where the weights of ``completely round'' and ``dust lane'' are equal, as shown by the light yellow boxes and dashed lines. An increase of about $+0.1$ weight in ``dust lane" feature beyond this boundary contributes to a relatively ideal D-ETGs, as shown by the dark yellow boxes. The farther the weight is from this boundary, the weaker the influence of the other feature in the generated images.}
    % \CHENRUI{1. Including this in the main text may occupy valuable space and potentially attract additional reviewers' comments, so I suggest to place it in the appendix. 2. If we plot weight vs. realism,  could the referee consider the analysis incomplete due to insufficient attention to diversity and consistency?}
    \label{fig:realism_for_prompt_weight_b}
\end{figure}

\section{Focal Loss Implementation}\label{appendix:focal_loss}

The focal loss is implemented by first computing the standard cross-entropy loss for each sample without reduction. 
Let $p_t$ denote the predicted probability of the ground-truth class, obtained as
\[
p_t = \exp(-\mathrm{CE}),
\]
where $\mathrm{CE}$ is the cross-entropy loss. 

The final focal loss for each sample is then calculated as
\[
\mathrm{FL} = \alpha (1 - p_t)^\gamma \cdot \mathrm{CE},
\]
where $\alpha$ is a weighting factor to address class imbalance, and $\gamma$ is the focusing parameter that reduces the relative loss contribution from well-classified examples, thus emphasizing hard-classified examples. Following the parameter settings reported in previous work by \cite{Liu2025addressing}, we adopted the same parameters with $\alpha = 0.25$ and $\gamma = 4$ to balance minority and majority classes while emphasizing hard-to-classify samples. After parameter adjustments on our dataset, this configuration consistently yielded stable convergence and superior validation performance compared with alternative parameter combinations, confirming its suitability for our task.

The per-sample focal losses are aggregated by taking the mean over the batch, which stabilizes training and ensures that the loss magnitude is independent of the batch size. In our implementation, we set the reduction mode to \texttt{mean} to achieve this effect.

\section{Additional Sanity Check on our D-ETG Samples}\label{appendix:sanity}

We here present the r-band magnitude and stellar mass distribution to complete our discussions in Section~\ref{subsec:few-shot}. Having identified 520 new D-ETGs using ML models enhanced with synthetic images, we compare their properties with the D-ETG sample from \cite{KAVIRAJ2012} to validate their shared physical consistency and with a control sample of early-type galaxies (ETGs) to ensure rigorous sample control.

\begin{figure}[H]
    \centering
    \gridline{
    \fig{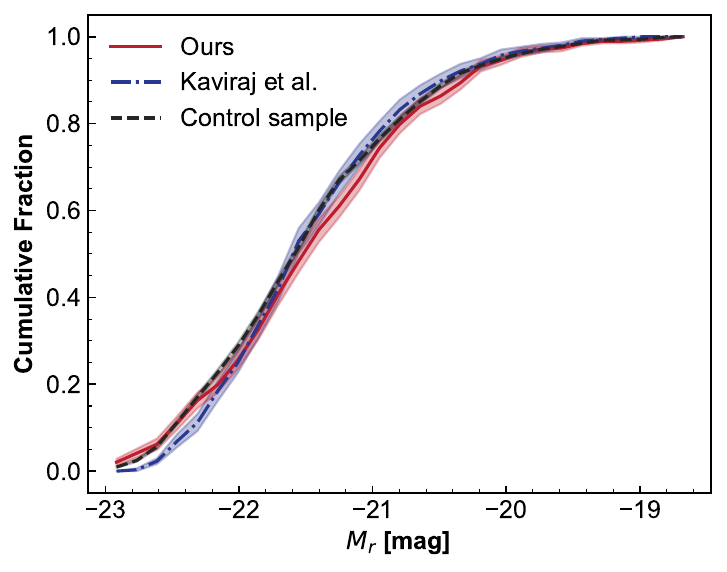}{0.48\textwidth}{(a) r-band Magnitude Comparison}
    \fig{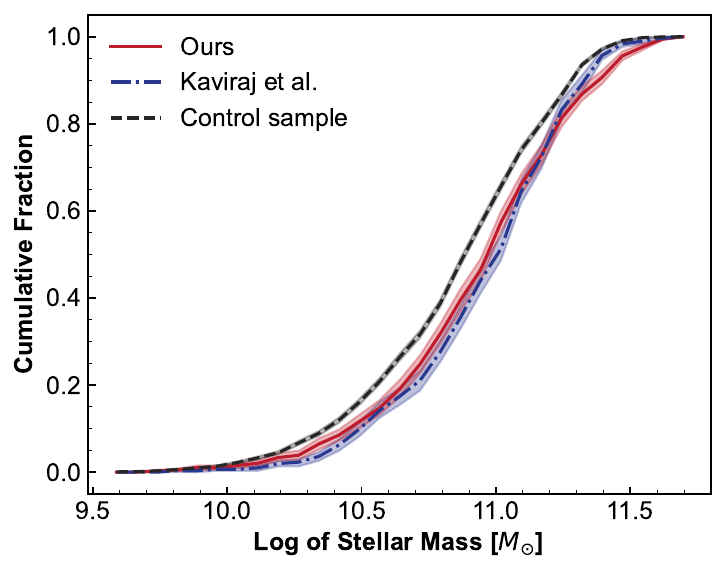}{0.48\textwidth}{(b) Stellar Mass Comparison}
    }
    \caption{Cumulative distribution functions (CDFs) comparing r-band magnitudes and stellar masses of our newly identified D-ETGs (red), the D-ETG sample from \cite{KAVIRAJ2012} (blue), and control ETGs (black). The close alignment between our sample and \cite{KAVIRAJ2012} confirms consistent physical properties for dusty early-type galaxies, while systematic differences from control ETGs validate the distinctiveness of the dusty morphology subset and the rigor of our sample control.}
    \label{fig:distributions}
\end{figure}

\section{Compared to GPT-4o}

As we have demonstrated that our fine-tuned diffusion model can generate high-quality galaxy images adhere to instruct we compare our diffusion model's results with the state-of-the-art industrial text-to-image model GPT-4o\footnote{\hyperlink{https://openai.com/index/introducing-4o-image-generation/}{https://openai.com/index/introducing-4o-image-generation/}} here, having demonstrated in the main text that our model generates high-fidelity, scientifically meaningful galaxy images. As shown in Figure~\ref{fig:gpt4o}, industrial text-to-image models effectively capture key visual features and morphological concepts of galaxy images but require further refinement to accurately replicate the systematic characteristics of real observational data. 

To more comprehensively compare images generated by our GalaxySD with those from GPT-4o, we conducted evaluations using the metrics introduced in the paper, namely realism and consistency. Because GPT-4o’s image generation function is usage-limited, each user could submit only a small number of text-to-image requests, making it infeasible to compute diversity in a statistically meaningful way. Consequently, our comparison with GPT-4o focuses primarily on realism and consistency, while the diversity metric is reported only for our model to highlight its ability to produce varied outputs under identical prompts. Figure~\ref{fig:realism_consistency_gpt4o} indicates the differences, our scores minus GPT-4o’s scores, on both axes. Using differences rather than absolute scores could directly visualize the performance gap and makes it clear whether our model performs better (positive values) or worse (negative values) than GPT-4o. Nearly all cases lie above or to the right of the gray dashed zero lines, with a few located directly on the lines, indicating that our GalaxySD generally outperforms GPT-4o in both realism and consistency.

\begin{figure}
    \centering
    \includegraphics[width=0.6\linewidth]{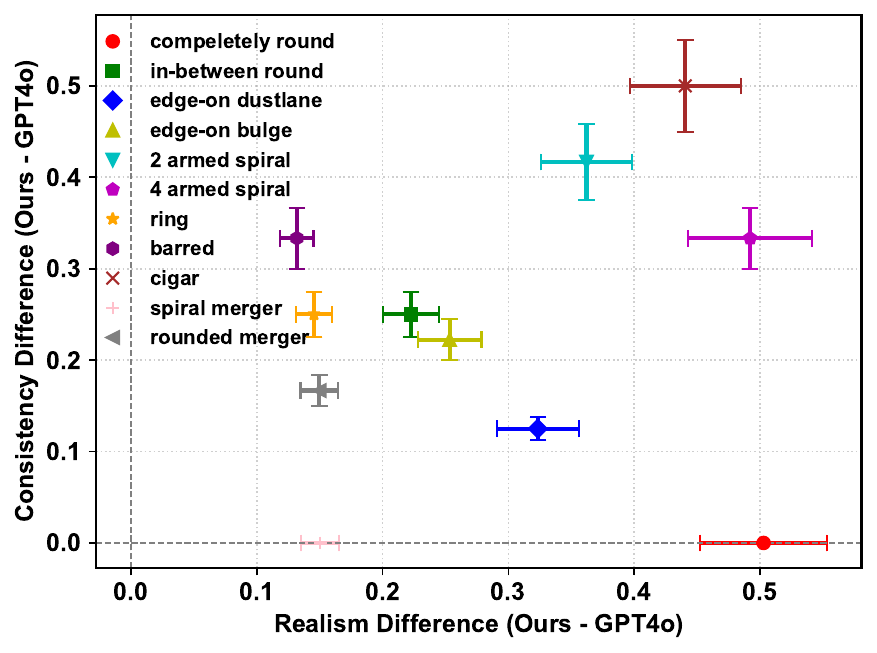}
    \caption{Scatter plot of the differences in realism and consistency scores between galaxy images generated by our GalaxySD and GPT-4o. The horizontal and vertical axes represent the differences, i.e., our scores minus GPT-4o’s scores. The error bars indicate ±10\% of the absolute value of each difference, reflecting the uncertainty or variability in the measurements due to the limited number of GPT-4o images. Different colors and markers distinguish the detailed morphological prompts.}
    \label{fig:realism_consistency_gpt4o}
\end{figure}

For future astrophysical AI development, leveraging pretrained industrial generative models as a foundation is more pragmatic. As illustrated in our study, starting from \texttt{Stable-Diffusion-v1-5} and training on only 27,910 high-quality annotated galaxy images yielded strong performance. Industrial models, having been trained on billions of high-quality image-text pairs, already encode the majority of general visual concepts, streamlining downstream domain adaptation for specialized astronomical applications.

\begin{figure}[H]
    \centering
    \includegraphics[width=1\linewidth]{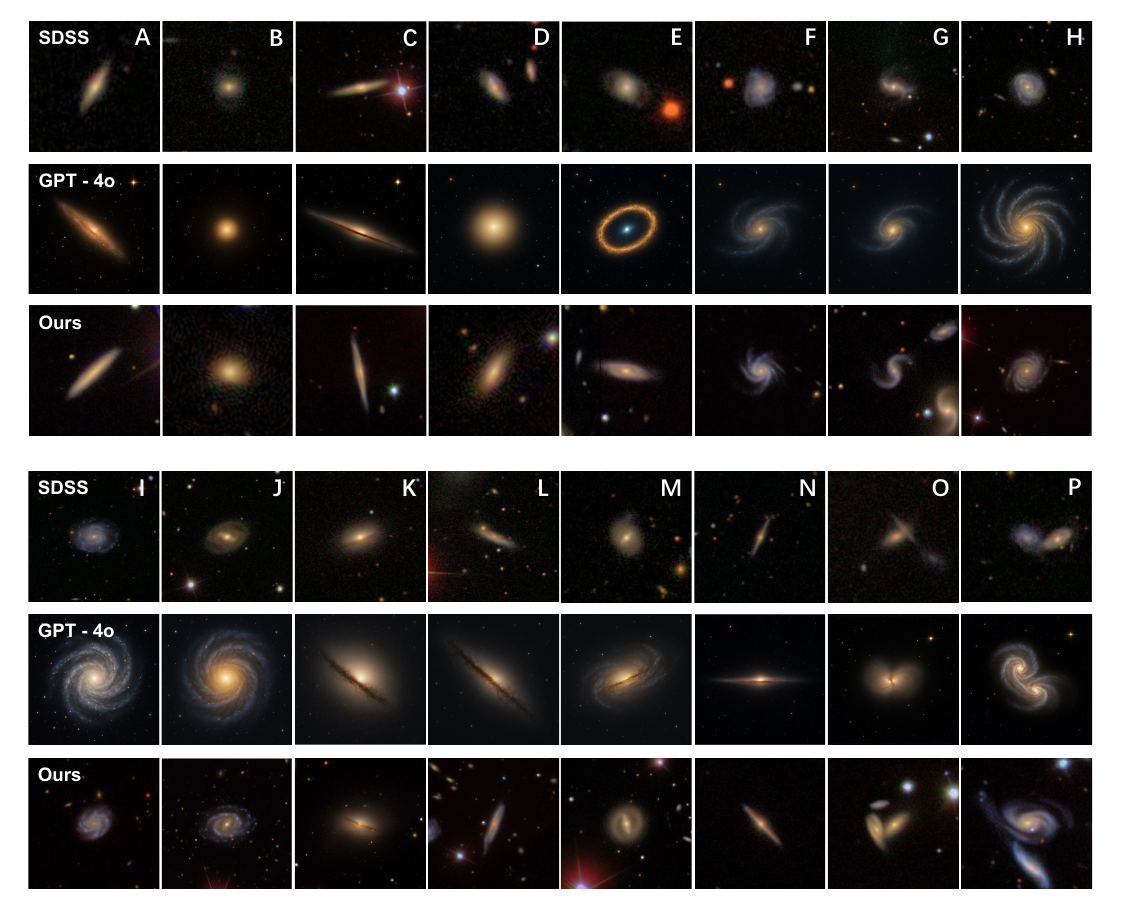}
    \caption{Comparison of galaxy images from SDSS, our model's simulations, and GPT-4o. Annotated capital letters correspond to generation prompts listed in Table~\ref{tab:prompts}. As demonstrated in our study, starting from industrial pretrained models (e.g., \texttt{Stable-Diffusion-v1-5}) and performing domain adaptation on science-specific datasets efficiently bridges the gap between general visual concepts and galaxy-specific observational features, enabling accurate reproduction of observational features critical for astrophysical applications.}
    \label{fig:gpt4o}
\end{figure}

%% For this sample we use BibTeX plus aasjournals.bst to generate the
%% the bibliography. The sample631.bib file was populated from ADS. To
%% get the citations to show in the compiled file do the following:
%%
%% pdflatex sample631.tex
%% bibtext sample631
%% pdflatex sample631.tex
%% pdflatex sample631.tex

\newpage

\bibliography{sample631}{}

\begin{thebibliography}{}
\expandafter\ifx\csname natexlab\endcsname\relax\def\natexlab#1{#1}\fi
\providecommand{\url}[1]{\href{#1}{#1}}
\providecommand{\dodoi}[1]{doi:~\href{http://doi.org/#1}{\nolinkurl{#1}}}
\providecommand{\doeprint}[1]{\href{http://ascl.net/#1}{\nolinkurl{http://ascl.net/#1}}}
\providecommand{\doarXiv}[1]{\href{https://arxiv.org/abs/#1}{\nolinkurl{https://arxiv.org/abs/#1}}}

\bibitem[{{Ahn} {et~al.}(2014){Ahn}, {Alexandroff}, {Allende Prieto}, {Anders}, {Anderson}, \& et~al.}]{AHNCHRISTOPHER2014}
{Ahn}, C.~P., {Alexandroff}, R., {Allende Prieto}, C., {et~al.} 2014, The Astrophysical Journal Supplement Series, 211, 17, \dodoi{10.1088/0067-0049/211/2/17}

\bibitem[{{Astolfi} {et~al.}(2024){Astolfi}, {Careil}, {Hall}, {Ma{\~n}as}, {Muckley}, {Verbeek}, {Romero Soriano}, \& {Drozdzal}}]{ASTOLFI2024}
{Astolfi}, P., {Careil}, M., {Hall}, M., {et~al.} 2024, arXiv e-prints, arXiv:2406.10429, \dodoi{10.48550/arXiv.2406.10429}

\bibitem[{Athanassoula \& Bosma(2019)}]{ATHANASSOULA2019}
Athanassoula, E., \& Bosma, A. 2019, Nature Astronomy, 3, 588, \dodoi{10.1038/s41550-019-0822-z}

\bibitem[{{Azizi} {et~al.}(2023){Azizi}, {Kornblith}, {Saharia}, {Norouzi}, \& {Fleet}}]{AZIZI2023}
{Azizi}, S., {Kornblith}, S., {Saharia}, C., {Norouzi}, M., \& {Fleet}, D.~J. 2023, arXiv e-prints, arXiv:2304.08466, \dodoi{10.48550/arXiv.2304.08466}

\bibitem[{{Ball} {et~al.}(2006){Ball}, {Brunner}, {Myers}, \& {Tcheng}}]{BALL2006}
{Ball}, N.~M., {Brunner}, R.~J., {Myers}, A.~D., \& {Tcheng}, D. 2006, The Astrophysical Journal, 650, 497, \dodoi{10.1086/507440}

\bibitem[{{Banerji} {et~al.}(2010){Banerji}, {Lahav}, {Lintott}, {Abdalla}, {Schawinski}, \& et~al.}]{BANERJI2010}
{Banerji}, M., {Lahav}, O., {Lintott}, C.~J., {et~al.} 2010, Monthly Notices of the Royal Astronomical Society, 406, 342, \dodoi{10.1111/j.1365-2966.2010.16713.x}

\bibitem[{{Banfield} {et~al.}(2015){Banfield}, {Wong}, {Willett}, {Norris}, {Rudnick}, \& et~al.}]{BANFIELD2015}
{Banfield}, J.~K., {Wong}, O.~I., {Willett}, K.~W., {et~al.} 2015, Monthly Notices of the Royal Astronomical Society, 453, 2326, \dodoi{10.1093/mnras/stv1688}

\bibitem[{{Bezanson} {et~al.}(2024){Bezanson}, {Labbe}, {Whitaker}, {Leja}, {Price}, \& et~al.}]{BEZANSON2024}
{Bezanson}, R., {Labbe}, I., {Whitaker}, K.~E., {et~al.} 2024, \apj, 974, 92, \dodoi{10.3847/1538-4357/ad66cf}

\bibitem[{{Bhaskara} {et~al.}(2024){Bhaskara}, {Georgakis}, {Nash}, {Cameron}, {Bowkett}, \& et~al.}]{BHASKARA2024}
{Bhaskara}, R., {Georgakis}, G., {Nash}, J., {et~al.} 2024, arXiv e-prints, arXiv:2401.12414, \dodoi{10.48550/arXiv.2401.12414}

\bibitem[{{Bowles} {et~al.}(2023){Bowles}, {Tang}, {Vardoulaki}, {Alexander}, {Luo}, \& et~al.}]{BOWLES2023}
{Bowles}, M., {Tang}, H., {Vardoulaki}, E., {et~al.} 2023, Monthly Notices of the Royal Astronomical Society, 522, 2584, \dodoi{10.1093/mnras/stad1021}

\bibitem[{{Brownstein} {et~al.}(2012){Brownstein}, {Bolton}, {Schlegel}, {Eisenstein}, {Kochanek}, \& et~al.}]{BROWNSTEIN2012}
{Brownstein}, J.~R., {Bolton}, A.~S., {Schlegel}, D.~J., {et~al.} 2012, The Astrophysical Journal, 744, 41, \dodoi{10.1088/0004-637X/744/1/41}

\bibitem[{{Bruzual} \& {Charlot}(2003)}]{BC032003}
{Bruzual}, G., \& {Charlot}, S. 2003, Monthly Notices of the Royal Astronomical Society, 344, 1000, \dodoi{10.1046/j.1365-8711.2003.06897.x}

\bibitem[{{Buzzo} {et~al.}(2025){Buzzo}, {Forbes}, {Jarrett}, {Marleau}, {Duc}, \& et~al.}]{BUZZOUDG2025}
{Buzzo}, M.~L., {Forbes}, D.~A., {Jarrett}, T.~H., {et~al.} 2025, Monthly Notices of the Royal Astronomical Society, 536, 2536, \dodoi{10.1093/mnras/stae2700}

\bibitem[{{Calzetti} {et~al.}(2000){Calzetti}, {Armus}, {Bohlin}, {Kinney}, {Koornneef}, \& et~al.}]{CALZETTI2000}
{Calzetti}, D., {Armus}, L., {Bohlin}, R.~C., {et~al.} 2000, The Astrophysical Journal, 533, 682, \dodoi{10.1086/308692}

\bibitem[{{Campagne}(2025)}]{CAMPAGNE2025}
{Campagne}, J.-E. 2025, Monthly Notices of the Royal Astronomical Society, 539, 3445, \dodoi{10.1093/mnras/staf533}

\bibitem[{{Casey} {et~al.}(2023){Casey}, {Kartaltepe}, {Drakos}, {Franco}, {Harish}, \& et~al.}]{CASEY2023}
{Casey}, C.~M., {Kartaltepe}, J.~S., {Drakos}, N.~E., {et~al.} 2023, \apj, 954, 31, \dodoi{10.3847/1538-4357/acc2bc}

\bibitem[{{Chen} {et~al.}(2023){Chen}, {Grasha}, {Battisti}, {Kewley}, {Madore}, {Seibert}, {Rich}, \& {Beaton}}]{QIANHUI2023}
{Chen}, Q.-H., {Grasha}, K., {Battisti}, A.~J., {et~al.} 2023, \mnras, 519, 4801, \dodoi{10.1093/mnras/stac3790}

\bibitem[{{Chen} {et~al.}(2020){Chen}, {Kornblith}, {Norouzi}, \& {Hinton}}]{CHEN2020}
{Chen}, T., {Kornblith}, S., {Norouzi}, M., \& {Hinton}, G. 2020, arXiv e-prints, arXiv:2002.05709, \dodoi{10.48550/arXiv.2002.05709}

\bibitem[{{Cho} {et~al.}(2023){Cho}, {Hu}, {Garg}, {Anderson}, {Krishna}, \& et~al.}]{CHOJAEMIN2023}
{Cho}, J., {Hu}, Y., {Garg}, R., {et~al.} 2023, arXiv e-prints, arXiv:2310.18235, \dodoi{10.48550/arXiv.2310.18235}

\bibitem[{{Conselice}(2003)}]{CONSELICE2003}
{Conselice}, C.~J. 2003, \apjs, 147, 1, \dodoi{10.1086/375001}

\bibitem[{{Conselice} {et~al.}(2000){Conselice}, {Bershady}, \& {Jangren}}]{CONSELICE2000}
{Conselice}, C.~J., {Bershady}, M.~A., \& {Jangren}, A. 2000, \apj, 529, 886, \dodoi{10.1086/308300}

\bibitem[{{Crain} {et~al.}(2015){Crain}, {Schaye}, {Bower}, {Furlong}, {Schaller}, \& et~al.}]{CRAIN2015}
{Crain}, R.~A., {Schaye}, J., {Bower}, R.~G., {et~al.} 2015, Monthly Notices of the Royal Astronomical Society, 450, 1937, \dodoi{10.1093/mnras/stv725}

\bibitem[{{Cui} {et~al.}(2024){Cui}, {Gu}, \& {Shi}}]{JIANTONGCUI2024}
{Cui}, J., {Gu}, Q., \& {Shi}, Y. 2024, Monthly Notices of the Royal Astronomical Society, 528, 2391, \dodoi{10.1093/mnras/stae156}

\bibitem[{{Dabbech} {et~al.}(2022){Dabbech}, {Terris}, {Jackson}, {Ramatsoku}, {Smirnov}, \& et~al.}]{DABBECH2022}
{Dabbech}, A., {Terris}, M., {Jackson}, A., {et~al.} 2022, The Astrophysical Journal, 939, L4, \dodoi{10.3847/2041-8213/ac98af}

\bibitem[{{Dalcanton} {et~al.}(2004){Dalcanton}, {Yoachim}, \& {Bernstein}}]{DALCANTON2004}
{Dalcanton}, J.~J., {Yoachim}, P., \& {Bernstein}, R.~A. 2004, The Astrophysical Journal, 608, 189, \dodoi{10.1086/386358}

\bibitem[{{Davis} {et~al.}(2015){Davis}, {Rowlands}, {Allison}, {Shabala}, {Ting}, \& et~al.}]{DAVIS2015}
{Davis}, T.~A., {Rowlands}, K., {Allison}, J.~R., {et~al.} 2015, Monthly Notices of the Royal Astronomical Society, 449, 3503, \dodoi{10.1093/mnras/stv597}

\bibitem[{{Deng} {et~al.}(2025){Deng}, {Shu}, {Wang}, {Li}, {Caminha}, \& et~al.}]{DENGLIMENG2025}
{Deng}, L., {Shu}, Y., {Wang}, L., {et~al.} 2025, The Astrophysical Journal, 982, L23, \dodoi{10.3847/2041-8213/adbae5}

\bibitem[{{Desmons} {et~al.}(2024){Desmons}, {Brough}, \& {Lanusse}}]{DESMONS2024}
{Desmons}, A., {Brough}, S., \& {Lanusse}, F. 2024, Monthly Notices of the Royal Astronomical Society, 531, 4070, \dodoi{10.1093/mnras/stae1402}

\bibitem[{{Dhariwal} \& {Nichol}(2021)}]{DHARIWAL2021}
{Dhariwal}, P., \& {Nichol}, A. 2021, arXiv e-prints, arXiv:2105.05233, \dodoi{10.48550/arXiv.2105.05233}

\bibitem[{{Duev} {et~al.}(2019){Duev}, {Mahabal}, {Ye}, {Tirumala}, {Belicki}, \& et~al.}]{DUEVDMITRY2019}
{Duev}, D.~A., {Mahabal}, A., {Ye}, Q., {et~al.} 2019, Monthly Notices of the Royal Astronomical Society, 486, 4158, \dodoi{10.1093/mnras/stz1096}

\bibitem[{{Euclid Collaboration} {et~al.}(2022){Euclid Collaboration}, {Scaramella}, {Amiaux}, {Mellier}, {Burigana}, \& et~al.}]{EUCLID2022}
{Euclid Collaboration}, {Scaramella}, R., {Amiaux}, J., {et~al.} 2022, Astronomy and Astrophysics, 662, A112, \dodoi{10.1051/0004-6361/202141938}

\bibitem[{{Euclid Collaboration} {et~al.}(2025){Euclid Collaboration}, {Siudek}, {Huertas-Company}, {Smith}, {Martinez-Solaeche}, \& et~al.}]{ASTROPT2025}
{Euclid Collaboration}, {Siudek}, M., {Huertas-Company}, M., {et~al.} 2025, arXiv e-prints, arXiv:2503.15312, \dodoi{10.48550/arXiv.2503.15312}

\bibitem[{{Finkelstein} {et~al.}(2025){Finkelstein}, {Bagley}, {Arrabal Haro}, {Dickinson}, {Ferguson}, \& et~al.}]{FINKELSTEIN2025}
{Finkelstein}, S.~L., {Bagley}, M.~B., {Arrabal Haro}, P., {et~al.} 2025, \apjl, 983, L4, \dodoi{10.3847/2041-8213/adbbd3}

\bibitem[{{Fontanot} {et~al.}(2011){Fontanot}, {De Lucia}, {Wilman}, \& {Monaco}}]{FONTANOT2011}
{Fontanot}, F., {De Lucia}, G., {Wilman}, D., \& {Monaco}, P. 2011, \mnras, 416, 409, \dodoi{10.1111/j.1365-2966.2011.19047.x}

\bibitem[{{Foyle} {et~al.}(2010){Foyle}, {Rix}, {Walter}, \& {Leroy}}]{FOYLE2010}
{Foyle}, K., {Rix}, H.~W., {Walter}, F., \& {Leroy}, A.~K. 2010, \apj, 725, 534, \dodoi{10.1088/0004-637X/725/1/534}

\bibitem[{{Fraser-McKelvie} {et~al.}(2019){Fraser-McKelvie}, {Merrifield}, {Arag{\'o}n-Salamanca}, {Peterken}, {Masters}, {Krawczyk}, {Andrews}, {Knapen}, {Kruk}, {Schaefer}, {Smethurst}, {Riffel}, {Brownstein}, \& {Drory}}]{FRASER2019}
{Fraser-McKelvie}, A., {Merrifield}, M., {Arag{\'o}n-Salamanca}, A., {et~al.} 2019, \mnras, 488, L6, \dodoi{10.1093/mnrasl/slz085}

\bibitem[{{Friedli} {et~al.}(1994){Friedli}, {Benz}, \& {Kennicutt}}]{FRIEDLI1994}
{Friedli}, D., {Benz}, W., \& {Kennicutt}, R. 1994, \apjl, 430, L105, \dodoi{10.1086/187449}

\bibitem[{{Fudamoto} {et~al.}(2022){Fudamoto}, {Inoue}, \& {Sugahara}}]{FUDAMOTO2022}
{Fudamoto}, Y., {Inoue}, A.~K., \& {Sugahara}, Y. 2022, The Astrophysical Journal, 938, L24, \dodoi{10.3847/2041-8213/ac982b}

\bibitem[{{Gardner} {et~al.}(2023){Gardner}, {Mather}, {Abbott}, {Abell}, {Abernathy}, \& et~al.}]{JWST2023}
{Gardner}, J.~P., {Mather}, J.~C., {Abbott}, R., {et~al.} 2023, Publications of the Astronomical Society of the Pacific, 135, 068001, \dodoi{10.1088/1538-3873/acd1b5}

\bibitem[{{Gupta} {et~al.}(2023){Gupta}, {Yu}, {Sohn}, {Gu}, {Hahn}, {Fei-Fei}, {Essa}, {Jiang}, \& {Lezama}}]{GUPTA2023}
{Gupta}, A., {Yu}, L., {Sohn}, K., {et~al.} 2023, arXiv e-prints, arXiv:2312.06662, \dodoi{10.48550/arXiv.2312.06662}

\bibitem[{{Hackstein} {et~al.}(2023){Hackstein}, {Kinakh}, {Bailer}, \& {Melchior}}]{HACKSTEIN2023}
{Hackstein}, S., {Kinakh}, V., {Bailer}, C., \& {Melchior}, M. 2023, Astronomy and Computing, 42, 100685, \dodoi{10.1016/j.ascom.2022.100685}

\bibitem[{{Hart} {et~al.}(2016){Hart}, {Bamford}, {Willett}, {Masters}, {Cardamone}, {Lintott}, {Mackay}, {Nichol}, {Rosslowe}, {Simmons}, \& {Smethurst}}]{gz2hart2016}
{Hart}, R.~E., {Bamford}, S.~P., {Willett}, K.~W., {et~al.} 2016, \mnras, 461, 3663, \dodoi{10.1093/mnras/stw1588}

\bibitem[{{Haslbauer} {et~al.}(2022){Haslbauer}, {Banik}, {Kroupa}, {Wittenburg}, \& {Javanmardi}}]{HASLBAUER2022}
{Haslbauer}, M., {Banik}, I., {Kroupa}, P., {Wittenburg}, N., \& {Javanmardi}, B. 2022, The Astrophysical Journal, 925, 183, \dodoi{10.3847/1538-4357/ac46ac}

\bibitem[{{Hausen} \& {Robertson}(2020)}]{HAUSEN2020}
{Hausen}, R., \& {Robertson}, B.~E. 2020, \apjs, 248, 20, \dodoi{10.3847/1538-4365/ab8868}

\bibitem[{{Hayat} {et~al.}(2021){Hayat}, {Stein}, {Harrington}, {Luki{\'c}}, \& {Mustafa}}]{HAYAT2021}
{Hayat}, M.~A., {Stein}, G., {Harrington}, P., {Luki{\'c}}, Z., \& {Mustafa}, M. 2021, The Astrophysical Journal, 911, L33, \dodoi{10.3847/2041-8213/abf2c7}

\bibitem[{{He} {et~al.}(2016){He}, {Zhang}, {Ren}, \& {Sun}}]{HE2016}
{He}, K., {Zhang}, X., {Ren}, S., \& {Sun}, J. 2016, in 2016 IEEE Conference on Computer Vision and Pattern Recognition (CVPR, 1, \dodoi{10.1109/CVPR.2016.90}

\bibitem[{{Ho} {et~al.}(2020){Ho}, {Jain}, \& {Abbeel}}]{HO2020}
{Ho}, J., {Jain}, A., \& {Abbeel}, P. 2020, arXiv e-prints, arXiv:2006.11239, \dodoi{10.48550/arXiv.2006.11239}

\bibitem[{{Ho} {et~al.}(2022){Ho}, {Chan}, {Saharia}, {Whang}, {Gao}, {Gritsenko}, {Kingma}, {Poole}, {Norouzi}, {Fleet}, \& {Salimans}}]{JONATHAN2022}
{Ho}, J., {Chan}, W., {Saharia}, C., {et~al.} 2022, arXiv e-prints, arXiv:2210.02303, \dodoi{10.48550/arXiv.2210.02303}

\bibitem[{{Hood} {et~al.}(2018){Hood}, {Kannappan}, {Stark}, {Dell'Antonio}, {Moffett}, \& et~al.}]{HOOD2018}
{Hood}, C.~E., {Kannappan}, S.~J., {Stark}, D.~V., {et~al.} 2018, The Astrophysical Journal, 857, 144, \dodoi{10.3847/1538-4357/aab719}

\bibitem[{{Hopkins} {et~al.}(2023){Hopkins}, {Gurvich}, {Shen}, {Hafen}, {Grudi{\'c}}, \& et~al.}]{HOPKINS2023}
{Hopkins}, P.~F., {Gurvich}, A.~B., {Shen}, X., {et~al.} 2023, Monthly Notices of the Royal Astronomical Society, 525, 2241, \dodoi{10.1093/mnras/stad1902}

\bibitem[{{Hopkins} {et~al.}(2010){Hopkins}, {Bundy}, {Croton}, {Hernquist}, {Keres}, {Khochfar}, {Stewart}, {Wetzel}, \& {Younger}}]{HOPKINS2010}
{Hopkins}, P.~F., {Bundy}, K., {Croton}, D., {et~al.} 2010, \apj, 715, 202, \dodoi{10.1088/0004-637X/715/1/202}

\bibitem[{{Hu} {et~al.}(2023){Hu}, {Liu}, {Kasai}, {Wang}, {Ostendorf}, \& et~al.}]{YUSHIHU2023}
{Hu}, Y., {Liu}, B., {Kasai}, J., {et~al.} 2023, arXiv e-prints, arXiv:2303.11897, \dodoi{10.48550/arXiv.2303.11897}

\bibitem[{{Irureta-Goyena} {et~al.}(2025){Irureta-Goyena}, {Rachith}, {Hellmich}, {Kneib}, {Altieri}, \& et~al.}]{IRURETAGOYENA2025}
{Irureta-Goyena}, B.~Y., {Rachith}, E., {Hellmich}, S., {et~al.} 2025, Astronomy and Astrophysics, 694, A49, \dodoi{10.1051/0004-6361/202452756}

\bibitem[{{Ivezi{\'c}} {et~al.}(2019){Ivezi{\'c}}, {Kahn}, {Tyson}, {Abel}, {Acosta}, \& et~al.}]{LSST2019}
{Ivezi{\'c}}, {\v{Z}}., {Kahn}, S.~M., {Tyson}, J.~A., {et~al.} 2019, The Astrophysical Journal, 873, 111, \dodoi{10.3847/1538-4357/ab042c}

\bibitem[{{Kaviraj} {et~al.}(2012){Kaviraj}, {Ting}, {Bureau}, {Shabala}, {Crockett}, \& et~al.}]{KAVIRAJ2012}
{Kaviraj}, S., {Ting}, Y.-S., {Bureau}, M., {et~al.} 2012, Monthly Notices of the Royal Astronomical Society, 423, 49, \dodoi{10.1111/j.1365-2966.2012.20957.x}

\bibitem[{{Keerthi-Vasan} {et~al.}(2023){Keerthi-Vasan}, {Sheng}, {Jones}, {Choi}, \& {Sharpnack}}]{KEERTHI2023}
{Keerthi-Vasan}, G.~C., {Sheng}, S., {Jones}, T., {Choi}, C.~P., \& {Sharpnack}, J. 2023, Monthly Notices of the Royal Astronomical Society, 524, 5368, \dodoi{10.1093/mnras/stad1709}

\bibitem[{{Khalid} {et~al.}(2024){Khalid}, {Brough}, {Martin}, {Kimmig}, {Lagos}, \& et~al.}]{KHALID2024}
{Khalid}, A., {Brough}, S., {Martin}, G., {et~al.} 2024, Monthly Notices of the Royal Astronomical Society, 530, 4422, \dodoi{10.1093/mnras/stae1064}

\bibitem[{{Kim} \& {Brunner}(2017)}]{KIMEDWARD2017}
{Kim}, E.~J., \& {Brunner}, R.~J. 2017, Monthly Notices of the Royal Astronomical Society, 464, 4463, \dodoi{10.1093/mnras/stw2672}

\bibitem[{Kingma \& Ba(2015)}]{Kingma2015adam}
Kingma, D.~P., \& Ba, J. 2015, in 3rd International Conference on Learning Representations, {ICLR} 2015, San Diego, CA, USA, May 7-9, 2015, Conference Track Proceedings, ed. Y.~Bengio \& Y.~LeCun.
\newblock \url{http://arxiv.org/abs/1412.6980}

\bibitem[{{Kingma} \& {Welling}(2013)}]{Kingma2013}
{Kingma}, D.~P., \& {Welling}, M. 2013, arXiv e-prints, arXiv:1312.6114, \dodoi{10.48550/arXiv.1312.6114}

\bibitem[{{Kohandel} {et~al.}(2024){Kohandel}, {Pallottini}, {Ferrara}, {Zanella}, {Rizzo}, \& {Carniani}}]{KOHANDEL2024}
{Kohandel}, M., {Pallottini}, A., {Ferrara}, A., {et~al.} 2024, \aap, 685, A72, \dodoi{10.1051/0004-6361/202348209}

\bibitem[{{Kreckel} {et~al.}(2013){Kreckel}, {Groves}, {Schinnerer}, {Johnson}, {Aniano}, \& et~al.}]{KRECKEL2013}
{Kreckel}, K., {Groves}, B., {Schinnerer}, E., {et~al.} 2013, The Astrophysical Journal, 771, 62, \dodoi{10.1088/0004-637X/771/1/62}

\bibitem[{{Lanusse} {et~al.}(2021){Lanusse}, {Mandelbaum}, {Ravanbakhsh}, {Li}, {Freeman}, \& et~al.}]{LANUSSE2021}
{Lanusse}, F., {Mandelbaum}, R., {Ravanbakhsh}, S., {et~al.} 2021, Monthly Notices of the Royal Astronomical Society, 504, 5543, \dodoi{10.1093/mnras/stab1214}

\bibitem[{{Leoni} {et~al.}(2022){Leoni}, {Ishida}, {Peloton}, \& {M{\"o}ller}}]{LEONIM2022}
{Leoni}, M., {Ishida}, E.~E.~O., {Peloton}, J., \& {M{\"o}ller}, A. 2022, Astronomy and Astrophysics, 663, A13, \dodoi{10.1051/0004-6361/202142715}

\bibitem[{{Li} {et~al.}(2023){Li}, {Greene}, {Greco}, {Huang}, {Melchior}, \& et~al.}]{JIAXUANLI2023}
{Li}, J., {Greene}, J.~E., {Greco}, J.~P., {et~al.} 2023, The Astrophysical Journal, 955, 1, \dodoi{10.3847/1538-4357/ace829}

\bibitem[{{Li} {et~al.}(2024){Li}, {Melchior}, {Hahn}, \& {Huang}}]{JIAXUAN2024}
{Li}, J., {Melchior}, P., {Hahn}, C., \& {Huang}, S. 2024, The Astronomical Journal, 167, 16, \dodoi{10.3847/1538-3881/ad0be4}

\bibitem[{{Li} {et~al.}(2020){Li}, {Napolitano}, {Tortora}, {Spiniello}, {Koopmans}, \& et~al.}]{LI2020}
{Li}, R., {Napolitano}, N.~R., {Tortora}, C., {et~al.} 2020, The Astrophysical Journal, 899, 30, \dodoi{10.3847/1538-4357/ab9dfa}

\bibitem[{{Lin} {et~al.}(2020){Lin}, {Li}, {Du}, {Wang}, {Xiao}, {Bureau}, {Fraser-McKelvie}, {Masters}, {Lin}, {Wake}, \& {Hao}}]{LIN2020}
{Lin}, L., {Li}, C., {Du}, C., {et~al.} 2020, \mnras, 499, 1406, \dodoi{10.1093/mnras/staa2913}

\bibitem[{{Lin} {et~al.}(2023{\natexlab{a}}){Lin}, {Liu}, {Li}, \& {Yang}}]{LINSHANCHUAN2023}
{Lin}, S., {Liu}, B., {Li}, J., \& {Yang}, X. 2023{\natexlab{a}}, arXiv e-prints, arXiv:2305.08891, \dodoi{10.48550/arXiv.2305.08891}

\bibitem[{Lin {et~al.}(2017)Lin, Goyal, Girshick, He, \& Doll{\'a}r}]{lin2017focal}
Lin, T.-Y., Goyal, P., Girshick, R., He, K., \& Doll{\'a}r, P. 2017, in Proceedings of the IEEE international conference on computer vision, 2980--2988

\bibitem[{{Lin} {et~al.}(2023{\natexlab{b}}){Lin}, {Cai}, {Zou}, {Li}, {Chen}, \& et~al.}]{XIAOJINGLIN2023}
{Lin}, X., {Cai}, Z., {Zou}, S., {et~al.} 2023{\natexlab{b}}, The Astrophysical Journal, 944, L59, \dodoi{10.3847/2041-8213/aca1c4}

\bibitem[{{Lintott} {et~al.}(2008){Lintott}, {Schawinski}, {Slosar}, {Land}, {Bamford}, {Thomas}, {Raddick}, {Nichol}, {Szalay}, {Andreescu}, {Murray}, \& {Vandenberg}}]{Lintott2008gz1sdss}
{Lintott}, C.~J., {Schawinski}, K., {Slosar}, A., {et~al.} 2008, \mnras, 389, 1179, \dodoi{10.1111/j.1365-2966.2008.13689.x}

\bibitem[{{Liu} {et~al.}(2025{\natexlab{a}}){Liu}, {Quan}, {Su}, {Guo}, {Liu}, \& et~al.}]{LIUTIE2025}
{Liu}, T., {Quan}, Y., {Su}, Y., {et~al.} 2025{\natexlab{a}}, arXiv e-prints, arXiv:2502.16807, \dodoi{10.48550/arXiv.2502.16807}

\bibitem[{{Liu} {et~al.}(2025{\natexlab{b}}){Liu}, {Jin}, {Zhao}, {Wang}, \& {Shen}}]{Liu2025addressing}
{Liu}, Y., {Jin}, J., {Zhao}, H., {Wang}, Z., \& {Shen}, Y. 2025{\natexlab{b}}, \apjs, 276, 39, \dodoi{10.3847/1538-4365/ad9dec}

\bibitem[{{Lizarraga} {et~al.}(2024){Lizarraga}, {Hanchen Jiang}, {Nowack}, {Li}, {Nian Wu}, \& et~al.}]{LIZARRAGA2024}
{Lizarraga}, A., {Hanchen Jiang}, E., {Nowack}, J., {et~al.} 2024, arXiv e-prints, arXiv:2411.18440, \dodoi{10.48550/arXiv.2411.18440}

\bibitem[{{Lotz} {et~al.}(2004){Lotz}, {Primack}, \& {Madau}}]{LOTZ2004}
{Lotz}, J.~M., {Primack}, J., \& {Madau}, P. 2004, The Astronomical Journal, 128, 163, \dodoi{10.1086/421849}

\bibitem[{{Ma} {et~al.}(2025){Ma}, {Huang}, {Yan}, {Chen}, {Duan}, \& et~al.}]{STEP2025}
{Ma}, G., {Huang}, H., {Yan}, K., {et~al.} 2025, arXiv e-prints, arXiv:2502.10248, \dodoi{10.48550/arXiv.2502.10248}

\bibitem[{{Masters} {et~al.}(2010){Masters}, {Mosleh}, {Romer}, {Nichol}, {Bamford}, \& et~al.}]{MASTERS2010}
{Masters}, K.~L., {Mosleh}, M., {Romer}, A.~K., {et~al.} 2010, Monthly Notices of the Royal Astronomical Society, 405, 783, \dodoi{10.1111/j.1365-2966.2010.16503.x}

\bibitem[{{Metcalf} {et~al.}(2019){Metcalf}, {Meneghetti}, {Avestruz}, {Bellagamba}, {Bom}, \& et~al.}]{METCALF2019}
{Metcalf}, R.~B., {Meneghetti}, M., {Avestruz}, C., {et~al.} 2019, Astronomy and Astrophysics, 625, A119, \dodoi{10.1051/0004-6361/201832797}

\bibitem[{{Mohale} \& {Lochner}(2024)}]{MOHALE2024}
{Mohale}, K., \& {Lochner}, M. 2024, Monthly Notices of the Royal Astronomical Society, 530, 1274, \dodoi{10.1093/mnras/stae926}

\bibitem[{{Morales} {et~al.}(2018){Morales}, {Mart{\'\i}nez-Delgado}, {Grebel}, {Cooper}, {Javanmardi}, \& et~al.}]{MORALES2018}
{Morales}, G., {Mart{\'\i}nez-Delgado}, D., {Grebel}, E.~K., {et~al.} 2018, Astronomy and Astrophysics, 614, A143, \dodoi{10.1051/0004-6361/201732271}

\bibitem[{{Nedkova} {et~al.}(2024){Nedkova}, {H{\"a}u{\ss}ler}, {Marchesini}, {Brammer}, {Feinstein}, {Johnston}, {Kartaltepe}, {Koekemoer}, {Martis}, {Muzzin}, {Rafelski}, {Shipley}, {Skelton}, {Stefanon}, {van der Wel}, \& {Whitaker}}]{NEDKOVA2024}
{Nedkova}, K.~V., {H{\"a}u{\ss}ler}, B., {Marchesini}, D., {et~al.} 2024, \mnras, 532, 3747, \dodoi{10.1093/mnras/stae1702}

\bibitem[{{Neeleman} {et~al.}(2020){Neeleman}, {Prochaska}, {Kanekar}, \& {Rafelski}}]{NEELEMAN2020}
{Neeleman}, M., {Prochaska}, J.~X., {Kanekar}, N., \& {Rafelski}, M. 2020, \nat, 581, 269, \dodoi{10.1038/s41586-020-2276-y}

\bibitem[{{Nelson} {et~al.}(2019){Nelson}, {Springel}, {Pillepich}, {Rodriguez-Gomez}, {Torrey}, \& et~al.}]{NELSON2019}
{Nelson}, D., {Springel}, V., {Pillepich}, A., {et~al.} 2019, Computational Astrophysics and Cosmology, 6, 2, \dodoi{10.1186/s40668-019-0028-x}

\bibitem[{{Nelson} {et~al.}(2023){Nelson}, {Suess}, {Bezanson}, {Price}, {van Dokkum}, {Leja}, {Wang}, {Whitaker}, {Labb{\'e}}, {Barrufet}, {Brammer}, {Eisenstein}, {Gibson}, {Hartley}, {Johnson}, {Heintz}, {Mathews}, {Miller}, {Oesch}, {Sandles}, {Setton}, {Speagle}, {Tacchella}, {Tadaki}, {{\"U}bler}, \& {Weaver}}]{NELSON2023}
{Nelson}, E.~J., {Suess}, K.~A., {Bezanson}, R., {et~al.} 2023, \apjl, 948, L18, \dodoi{10.3847/2041-8213/acc1e1}

\bibitem[{{Noll} {et~al.}(2009){Noll}, {Burgarella}, {Giovannoli}, {Buat}, {Marcillac}, \& et~al.}]{CIGALE2009}
{Noll}, S., {Burgarella}, D., {Giovannoli}, E., {et~al.} 2009, Astronomy and Astrophysics, 507, 1793, \dodoi{10.1051/0004-6361/200912497}

\bibitem[{{O'Briain} {et~al.}(2021){O'Briain}, {Ting}, {Fabbro}, {Yi}, {Venn}, \& {Bialek}}]{OBRIAIN2021}
{O'Briain}, T., {Ting}, Y.-S., {Fabbro}, S., {et~al.} 2021, \apj, 906, 130, \dodoi{10.3847/1538-4357/abca96}

\bibitem[{{Omori} {et~al.}(2023){Omori}, {Bottrell}, {Walmsley}, {Yesuf}, {Goulding}, \& et~al.}]{OMORI2023}
{Omori}, K.~C., {Bottrell}, C., {Walmsley}, M., {et~al.} 2023, Astronomy and Astrophysics, 679, A142, \dodoi{10.1051/0004-6361/202346743}

\bibitem[{{Ono} {et~al.}(2024){Ono}, {Park}, {Mudur}, {Ni}, {Cuesta-Lazaro}, \& et~al.}]{ONO2024}
{Ono}, V., {Park}, C.~F., {Mudur}, N., {et~al.} 2024, \apj, 970, 174, \dodoi{10.3847/1538-4357/ad5957}

\bibitem[{{O'Riordan} {et~al.}(2025){O'Riordan}, {Oldham}, {Nersesian}, {Li}, {Collett}, \& et~al.}]{ORIORDAN2025}
{O'Riordan}, C.~M., {Oldham}, L.~J., {Nersesian}, A., {et~al.} 2025, Astronomy and Astrophysics, 694, A145, \dodoi{10.1051/0004-6361/202453014}

\bibitem[{{Parker} {et~al.}(2024){Parker}, {Lanusse}, {Golkar}, {Sarra}, {Cranmer}, \& et~al.}]{ASTROCLIP2024}
{Parker}, L., {Lanusse}, F., {Golkar}, S., {et~al.} 2024, \mnras, 531, 4990, \dodoi{10.1093/mnras/stae1450}

\bibitem[{{Pearce-Casey} {et~al.}(2024){Pearce-Casey}, {Nagam}, {Wilde}, {Busillo}, {Ulivi}, \& et~al.}]{PEARCE2024}
{Pearce-Casey}, R., {Nagam}, B.~C., {Wilde}, J., {et~al.} 2024, arXiv e-prints, arXiv:2411.16808, \dodoi{10.48550/arXiv.2411.16808}

\bibitem[{Pearl(2009)}]{PEARL2009}
Pearl, J. 2009, Causality (Cambridge: Cambridge University Press)

\bibitem[{Pearl(2010)}]{PERAL2010}
---. 2010, International Journal of Biostatistics, 6, Article 7, \dodoi{10.2202/1557-4679.1203}

\bibitem[{{Pearson} {et~al.}(2024){Pearson}, {Santos}, {Goto}, {Huang}, {Kim}, {Matsuhara}, {Pollo}, {Ho}, {Hwang}, {Ma{\l}ek}, {Nakagawa}, {Romano}, {Serjeant}, {Suelves}, {Shim}, \& {White}}]{PEARSON2024}
{Pearson}, W.~J., {Santos}, D.~J.~D., {Goto}, T., {et~al.} 2024, \aap, 686, A94, \dodoi{10.1051/0004-6361/202349034}

\bibitem[{{Planck Collaboration} {et~al.}(2020){Planck Collaboration}, {Aghanim}, {Akrami}, {Ashdown}, {Aumont}, \& et~al.}]{PLANCK2020}
{Planck Collaboration}, {Aghanim}, N., {Akrami}, Y., {et~al.} 2020, \aap, 641, A6, \dodoi{10.1051/0004-6361/201833910}

\bibitem[{{Robertson} {et~al.}(2023){Robertson}, {Tacchella}, {Johnson}, {Hausen}, {Alabi}, \& et~al.}]{BRANT2023}
{Robertson}, B.~E., {Tacchella}, S., {Johnson}, B.~D., {et~al.} 2023, The Astrophysical Journal, 942, L42, \dodoi{10.3847/2041-8213/aca086}

\bibitem[{{Rombach} {et~al.}(2021){Rombach}, {Blattmann}, {Lorenz}, {Esser}, \& {Ommer}}]{ROMBACH2021}
{Rombach}, R., {Blattmann}, A., {Lorenz}, D., {Esser}, P., \& {Ommer}, B. 2021, arXiv e-prints, arXiv:2112.10752, \dodoi{10.48550/arXiv.2112.10752}

\bibitem[{{Ronneberger} {et~al.}(2015){Ronneberger}, {Fischer}, \& {Brox}}]{UNET2015}
{Ronneberger}, O., {Fischer}, P., \& {Brox}, T. 2015, arXiv e-prints, arXiv:1505.04597, \dodoi{10.48550/arXiv.1505.04597}

\bibitem[{{Rouhiainen} {et~al.}(2024){Rouhiainen}, {M{\"u}nchmeyer}, {Shiu}, {Gira}, \& {Lee}}]{ROUHIANINEN2024}
{Rouhiainen}, A., {M{\"u}nchmeyer}, M., {Shiu}, G., {Gira}, M., \& {Lee}, K. 2024, Physical Review D, 109, 123536, \dodoi{10.1103/PhysRevD.109.123536}

\bibitem[{{Rowe} {et~al.}(2015){Rowe}, {Jarvis}, {Mandelbaum}, {Bernstein}, {Bosch}, \& et~al.}]{GALSIM2015}
{Rowe}, B.~T.~P., {Jarvis}, M., {Mandelbaum}, R., {et~al.} 2015, Astronomy and Computing, 10, 121, \dodoi{10.1016/j.ascom.2015.02.002}

\bibitem[{{Rutherford} {et~al.}(2024){Rutherford}, {van de Sande}, {Croom}, {Valenzuela}, {Remus}, \& et~al.}]{RUTHERFORD2024}
{Rutherford}, T.~H., {van de Sande}, J., {Croom}, S.~M., {et~al.} 2024, Monthly Notices of the Royal Astronomical Society, 529, 810, \dodoi{10.1093/mnras/stae398}

\bibitem[{{Salim} {et~al.}(2016){Salim}, {Lee}, {Janowiecki}, {da Cunha}, {Dickinson}, \& et~al.}]{SALIM2016}
{Salim}, S., {Lee}, J.~C., {Janowiecki}, S., {et~al.} 2016, The Astrophysical Journal Supplement Series, 227, 2, \dodoi{10.3847/0067-0049/227/1/2}

\bibitem[{{Schinnerer} {et~al.}(2017){Schinnerer}, {Meidt}, {Colombo}, {Chandar}, {Dobbs}, {Garc{\'\i}a-Burillo}, {Hughes}, {Leroy}, {Pety}, {Querejeta}, {Kramer}, \& {Schuster}}]{SCHINNERER2017}
{Schinnerer}, E., {Meidt}, S.~E., {Colombo}, D., {et~al.} 2017, \apj, 836, 62, \dodoi{10.3847/1538-4357/836/1/62}

\bibitem[{{Sether} {et~al.}(2024){Sether}, {Giusarma}, \& {Reyes-Hurtado}}]{SETHER2024}
{Sether}, T., {Giusarma}, E., \& {Reyes-Hurtado}, M. 2024, arXiv e-prints, arXiv:2412.05131, \dodoi{10.48550/arXiv.2412.05131}

\bibitem[{{Shabala} {et~al.}(2012){Shabala}, {Ting}, {Kaviraj}, {Lintott}, {Crockett}, \& et~al.}]{SHABALA2012}
{Shabala}, S.~S., {Ting}, Y.-S., {Kaviraj}, S., {et~al.} 2012, Monthly Notices of the Royal Astronomical Society, 423, 59, \dodoi{10.1111/j.1365-2966.2012.20598.x}

\bibitem[{{Slijepcevic} {et~al.}(2022){Slijepcevic}, {Scaife}, {Walmsley}, {Bowles}, {Wong}, \& et~al.}]{INIGOV2022}
{Slijepcevic}, I.~V., {Scaife}, A. M.~M., {Walmsley}, M., {et~al.} 2022, Monthly Notices of the Royal Astronomical Society, 514, 2599, \dodoi{10.1093/mnras/stac1135}

\bibitem[{{Smith} {et~al.}(2022){Smith}, {Geach}, {Jackson}, {Arora}, {Stone}, \& et~al.}]{SMITH2022}
{Smith}, M.~J., {Geach}, J.~E., {Jackson}, R.~A., {et~al.} 2022, Monthly Notices of the Royal Astronomical Society, 511, 1808, \dodoi{10.1093/mnras/stac130}

\bibitem[{{Smith} {et~al.}(2024){Smith}, {Roberts}, {Angeloudi}, \& {Huertas-Company}}]{ASTROPT2024}
{Smith}, M.~J., {Roberts}, R.~J., {Angeloudi}, E., \& {Huertas-Company}, M. 2024, arXiv e-prints, arXiv:2405.14930, \dodoi{10.48550/arXiv.2405.14930}

\bibitem[{{Song} {et~al.}(2020){Song}, {Sohl-Dickstein}, {Kingma}, {Kumar}, {Ermon}, \& et~al.}]{SONGYANG2020}
{Song}, Y., {Sohl-Dickstein}, J., {Kingma}, D.~P., {et~al.} 2020, arXiv e-prints, arXiv:2011.13456, \dodoi{10.48550/arXiv.2011.13456}

\bibitem[{{Song} {et~al.}(2023){Song}, {He}, {Li}, {Ma}, {Ming}, {Mao}, {Pei}, {Peng}, {Hu}, {Yao}, \& {Zhang}}]{SONGZHIHANG2023}
{Song}, Z., {He}, Z., {Li}, X., {et~al.} 2023, arXiv e-prints, arXiv:2304.12205, \dodoi{10.48550/arXiv.2304.12205}

\bibitem[{{Speagle} {et~al.}(2019{\natexlab{a}}){Speagle}, {Leauthaud}, {Huang}, {Bradshaw}, {Ardila}, \& et~al.}]{JOSHUA2019}
{Speagle}, J.~S., {Leauthaud}, A., {Huang}, S., {et~al.} 2019{\natexlab{a}}, \mnras, 490, 5658, \dodoi{10.1093/mnras/stz2968}

\bibitem[{{Speagle} {et~al.}(2019{\natexlab{b}}){Speagle}, {Leauthaud}, {Huang}, {Bradshaw}, {Ardila}, {Capak}, {Eisenstein}, {Masters}, {Mandelbaum}, {More}, {Simet}, \& {Sif{\'o}n}}]{SPEAGLE2019}
---. 2019{\natexlab{b}}, \mnras, 490, 5658, \dodoi{10.1093/mnras/stz2968}

\bibitem[{{Stein} {et~al.}(2022){Stein}, {Blaum}, {Harrington}, {Medan}, \& {Luki{\'c}}}]{STEIN2022}
{Stein}, G., {Blaum}, J., {Harrington}, P., {Medan}, T., \& {Luki{\'c}}, Z. 2022, The Astrophysical Journal, 932, 107, \dodoi{10.3847/1538-4357/ac6d63}

\bibitem[{{Sun} {et~al.}(2024{\natexlab{a}}){Sun}, {Calzetti}, \& {Battisti}}]{BINGQING2024}
{Sun}, B., {Calzetti}, D., \& {Battisti}, A.~J. 2024{\natexlab{a}}, \apj, 973, 137, \dodoi{10.3847/1538-4357/ad6157}

\bibitem[{{Sun} {et~al.}(2023{\natexlab{a}}){Sun}, {Speagle}, {Huang}, {Ting}, \& {Cai}}]{ZEPHYR2023}
{Sun}, Z., {Speagle}, J.~S., {Huang}, S., {Ting}, Y.-S., \& {Cai}, Z. 2023{\natexlab{a}}, arXiv e-prints, arXiv:2310.20125, \dodoi{10.48550/arXiv.2310.20125}

\bibitem[{{Sun} {et~al.}(2023{\natexlab{b}}){Sun}, {Ting}, \& {Cai}}]{SUN2023}
{Sun}, Z., {Ting}, Y.-S., \& {Cai}, Z. 2023{\natexlab{b}}, The Astrophysical Journal Supplement Series, 269, 4, \dodoi{10.3847/1538-4365/acf2f1}

\bibitem[{{Sun} {et~al.}(2024{\natexlab{b}}){Sun}, {Ting}, {Liang}, {Duan}, {Huang}, \& et~al.}]{SUNZECHANG2024}
{Sun}, Z., {Ting}, Y.-S., {Liang}, Y., {et~al.} 2024{\natexlab{b}}, arXiv e-prints, arXiv:2409.14807, \dodoi{10.48550/arXiv.2409.14807}

\bibitem[{{Sweere} {et~al.}(2022){Sweere}, {Valtchanov}, {Lieu}, {Vojtekova}, {Verdugo}, \& et~al.}]{SWEERE2022}
{Sweere}, S.~F., {Valtchanov}, I., {Lieu}, M., {et~al.} 2022, Monthly Notices of the Royal Astronomical Society, 517, 4054, \dodoi{10.1093/mnras/stac2437}

\bibitem[{{Tang} {et~al.}(2020){Tang}, {Scaife}, {Wong}, {Kapi{\'n}ska}, {Rudnick}, \& et~al.}]{TANG2020}
{Tang}, H., {Scaife}, A.~M.~M., {Wong}, O.~I., {et~al.} 2020, Monthly Notices of the Royal Astronomical Society, 499, 68, \dodoi{10.1093/mnras/staa2805}

\bibitem[{{Tanoglidis} {et~al.}(2022){Tanoglidis}, {{\'C}iprijanovi{\'c}}, {Drlica-Wagner}, {Nord}, {Wang}, \& et~al.}]{TANOGLIDIS2022}
{Tanoglidis}, D., {{\'C}iprijanovi{\'c}}, A., {Drlica-Wagner}, A., {et~al.} 2022, Astronomy and Computing, 39, 100580, \dodoi{10.1016/j.ascom.2022.100580}

\bibitem[{{Tenachi} {et~al.}(2023){Tenachi}, {Ibata}, \& {Diakogiannis}}]{TENACHI2023}
{Tenachi}, W., {Ibata}, R., \& {Diakogiannis}, F.~I. 2023, The Astrophysical Journal, 959, 99, \dodoi{10.3847/1538-4357/ad014c}

\bibitem[{{Terris} {et~al.}(2023){Terris}, {Dabbech}, {Tang}, \& {Wiaux}}]{TERRIS2023}
{Terris}, M., {Dabbech}, A., {Tang}, C., \& {Wiaux}, Y. 2023, Monthly Notices of the Royal Astronomical Society, 518, 604, \dodoi{10.1093/mnras/stac2672}

\bibitem[{{Thuruthipilly} {et~al.}(2025){Thuruthipilly}, {Junais}, {Koda}, {Pollo}, {Yagi}, \& et~al.}]{THURUTHIPILLY2025}
{Thuruthipilly}, H., {Junais}, {Koda}, J., {et~al.} 2025, arXiv e-prints, arXiv:2502.03142, \dodoi{10.48550/arXiv.2502.03142}

\bibitem[{{Ting} {et~al.}(2019){Ting}, {Conroy}, {Rix}, \& {Cargile}}]{TING2019}
{Ting}, Y.-S., {Conroy}, C., {Rix}, H.-W., \& {Cargile}, P. 2019, The Astrophysical Journal, 879, 69, \dodoi{10.3847/1538-4357/ab2331}

\bibitem[{{Trussler} {et~al.}(2023){Trussler}, {Conselice}, {Adams}, {Maiolino}, {Nakajima}, \& et~al.}]{TRUSSLER2023}
{Trussler}, J. A.~A., {Conselice}, C.~J., {Adams}, N.~J., {et~al.} 2023, Monthly Notices of the Royal Astronomical Society, 525, 5328, \dodoi{10.1093/mnras/stad2553}

\bibitem[{Vaswani {et~al.}(2017)Vaswani, Shazeer, Parmar, Uszkoreit, Jones, Gomez, Kaiser, \& Polosukhin}]{vaswani2017attention}
Vaswani, A., Shazeer, N., Parmar, N., {et~al.} 2017, Advances in neural information processing systems, 30

\bibitem[{{Vavilova} {et~al.}(2021){Vavilova}, {Dobrycheva}, {Vasylenko}, {Elyiv}, {Melnyk}, \& et~al.}]{VAVILOVA2021}
{Vavilova}, I.~B., {Dobrycheva}, D.~V., {Vasylenko}, M.~Y., {et~al.} 2021, Astronomy and Astrophysics, 648, A122, \dodoi{10.1051/0004-6361/202038981}

\bibitem[{{Villaescusa-Navarro} {et~al.}(2021){Villaescusa-Navarro}, {Angl{\'e}s-Alc{\'a}zar}, {Genel}, {Spergel}, {Somerville}, \& et~al.}]{CAMELS2021}
{Villaescusa-Navarro}, F., {Angl{\'e}s-Alc{\'a}zar}, D., {Genel}, S., {et~al.} 2021, \apj, 915, 71, \dodoi{10.3847/1538-4357/abf7ba}

\bibitem[{{Vi{\v{c}}{\'a}nek Mart{\'\i}nez} {et~al.}(2024){Vi{\v{c}}{\'a}nek Mart{\'\i}nez}, {Baron Perez}, \& {Br{\"u}ggen}}]{VIVCANEK2024}
{Vi{\v{c}}{\'a}nek Mart{\'\i}nez}, T., {Baron Perez}, N., \& {Br{\"u}ggen}, M. 2024, Astronomy and Astrophysics, 691, A360, \dodoi{10.1051/0004-6361/202451429}

\bibitem[{{Walmsley} {et~al.}(2023{\natexlab{a}}){Walmsley}, {G{\'e}ron}, {Kruk}, {Scaife}, {Lintott}, \& et~al.}]{WALMSLEY2023}
{Walmsley}, M., {G{\'e}ron}, T., {Kruk}, S., {et~al.} 2023{\natexlab{a}}, Monthly Notices of the Royal Astronomical Society, 526, 4768, \dodoi{10.1093/mnras/stad2919}

\bibitem[{{Walmsley} {et~al.}(2022){Walmsley}, {Slijepcevic}, {Bowles}, \& {Scaife}}]{WALMSLEYMIKE2022}
{Walmsley}, M., {Slijepcevic}, I., {Bowles}, M.~R., \& {Scaife}, A. 2022, in Machine Learning for Astrophysics, 29, \dodoi{10.48550/arXiv.2206.11927}

\bibitem[{{Walmsley} {et~al.}(2023{\natexlab{b}}){Walmsley}, {G{\'e}ron}, {Kruk}, {Scaife}, {Lintott}, {Masters}, {Dawson}, {Dickinson}, {Fortson}, {Garland}, {Mantha}, {O'Ryan}, {Popp}, {Simmons}, {Baeten}, \& {Macmillan}}]{gzdesi2023}
---. 2023{\natexlab{b}}, \mnras, 526, 4768, \dodoi{10.1093/mnras/stad2919}

\bibitem[{{Wang} {et~al.}(2022){Wang}, {Zou}, {Cai}, {Prochaska}, {Sun}, \& et~al.}]{BEN2022}
{Wang}, B., {Zou}, J., {Cai}, Z., {et~al.} 2022, The Astrophysical Journal Supplement Series, 259, 28, \dodoi{10.3847/1538-4365/ac4504}

\bibitem[{{Wei} {et~al.}(2025){Wei}, {Huang}, {Li}, {Sun}, {Li}, \& et~al.}]{LEYAO2025}
{Wei}, L., {Huang}, S., {Li}, J., {et~al.} 2025, arXiv e-prints, arXiv:2505.14073, \dodoi{10.48550/arXiv.2505.14073}

\bibitem[{{Willett} {et~al.}(2013){Willett}, {Lintott}, {Bamford}, {Masters}, {Simmons}, {Casteels}, {Edmondson}, {Fortson}, {Kaviraj}, {Keel}, {Melvin}, {Nichol}, {Raddick}, {Schawinski}, {Simpson}, {Skibba}, {Smith}, \& {Thomas}}]{Willett2013gz2sdss}
{Willett}, K.~W., {Lintott}, C.~J., {Bamford}, S.~P., {et~al.} 2013, \mnras, 435, 2835, \dodoi{10.1093/mnras/stt1458}

\bibitem[{{Willett} {et~al.}(2017){Willett}, {Galloway}, {Bamford}, {Lintott}, {Masters}, {Scarlata}, {Simmons}, {Beck}, {Cardamone}, {Cheung}, {Edmondson}, {Fortson}, {Griffith}, {H{\"a}u{\ss}ler}, {Han}, {Hart}, {Melvin}, {Parrish}, {Schawinski}, {Smethurst}, \& {Smith}}]{gzhst2017}
{Willett}, K.~W., {Galloway}, M.~A., {Bamford}, S.~P., {et~al.} 2017, \mnras, 464, 4176, \dodoi{10.1093/mnras/stw2568}

\bibitem[{{Wittenburg} {et~al.}(2023){Wittenburg}, {Kroupa}, {Banik}, {Candlish}, \& {Samaras}}]{WITTENBURG2023}
{Wittenburg}, N., {Kroupa}, P., {Banik}, I., {Candlish}, G., \& {Samaras}, N. 2023, Monthly Notices of the Royal Astronomical Society, 523, 453, \dodoi{10.1093/mnras/stad1371}

\bibitem[{{Xie} {et~al.}(2021){Xie}, {Zhang}, {Cao}, {Lin}, {Bao}, {Yao}, {Dai}, \& {Hu}}]{ZHENDAXIE2021}
{Xie}, Z., {Zhang}, Z., {Cao}, Y., {et~al.} 2021, arXiv e-prints, arXiv:2111.09886, \dodoi{10.48550/arXiv.2111.09886}

\bibitem[{{Xu} {et~al.}(2023){Xu}, {McCully}, {Dong}, {Howell}, \& {Sen}}]{XUCHENGYUAN2023}
{Xu}, C., {McCully}, C., {Dong}, B., {Howell}, D.~A., \& {Sen}, P. 2023, The Astrophysical Journal, 942, 73, \dodoi{10.3847/1538-4357/ac9d91}

\bibitem[{{Yan} {et~al.}(2024){Yan}, {Sun}, \& {Ling}}]{YANHAOJING2024}
{Yan}, H., {Sun}, B., \& {Ling}, C. 2024, \apj, 975, 44, \dodoi{10.3847/1538-4357/ad7de9}

\bibitem[{{Yang} {et~al.}(2012){Yang}, {Mo}, {van den Bosch}, {Zhang}, \& {Han}}]{XIAOHUYANG2012}
{Yang}, X., {Mo}, H.~J., {van den Bosch}, F.~C., {Zhang}, Y., \& {Han}, J. 2012, The Astrophysical Journal, 752, 41, \dodoi{10.1088/0004-637X/752/1/41}

\bibitem[{{Zhang} \& {Brandt}(2021)}]{HENGYUE2021}
{Zhang}, H., \& {Brandt}, T.~D. 2021, The Astronomical Journal, 162, 139, \dodoi{10.3847/1538-3881/ac1348}

\bibitem[{{Zhang} \& {Bloom}(2020)}]{KEMING2020}
{Zhang}, K., \& {Bloom}, J.~S. 2020, The Astrophysical Journal, 889, 24, \dodoi{10.3847/1538-4357/ab3fa6}

\bibitem[{{Zhang} {et~al.}(2024){Zhang}, {Yin}, {Li}, \& {Xie}}]{PENGZEZHANG2024}
{Zhang}, P., {Yin}, H., {Li}, C., \& {Xie}, X. 2024, arXiv e-prints, arXiv:2403.08381, \dodoi{10.48550/arXiv.2403.08381}

\bibitem[{Zhao {et~al.}(2023)Zhao, Ting, Diao, \& Mao}]{XIAOSHENG2023}
Zhao, X., Ting, Y.-S., Diao, K., \& Mao, Y. 2023, Monthly Notices of the Royal Astronomical Society, 526, 1699, \dodoi{10.1093/mnras/stad2778}

\bibitem[{{Zhu} {et~al.}(2023){Zhu}, {Zhao}, {He}, {Zhong}, {Zhang}, {Guo}, {Chen}, \& {Zhang}}]{ZHUZHENGBANG2023}
{Zhu}, Z., {Zhao}, H., {He}, H., {et~al.} 2023, arXiv e-prints, arXiv:2311.01223, \dodoi{10.48550/arXiv.2311.01223}

\end{thebibliography}
\bibliographystyle{aasjournal}

%% This command is needed to show the entire author+affiliation list when
%% the collaboration and author truncation commands are used.  It has to
%% go at the end of the manuscript.
%\allauthors

%% Include this line if you are using the \added, \replaced, \deleted
%% commands to see a summary list of all changes at the end of the article.
%\listofchanges

\end{document}